\documentclass[aps,prd,showpacs,floatfix,preprintnumbers,nofootinbib,
amssymb,amsmath]{revtex4}
\usepackage{subfigure}
\usepackage{graphicx,bm,color}
\usepackage{amssymb}

\begin{document}
\title{Reparametrizing the
Polyakov$-$Nambu$-$Jona-Lasinio model}
\author{Abhijit Bhattacharyya}
\email{abhattacharyyacu@gmail.com}
\affiliation{Department of Physics, University of Calcutta,
             92, A.P.C Road, Kolkata-700009, INDIA}
\author{Sanjay K. Ghosh}
\email{sanjay@jcbose.ac.in}
\author{Soumitra Maity}
\email{soumitra.maity1984@gmail.com}
\author{Sibaji Raha}
\email{sibaji@jcbose.ac.in}
\author{Rajarshi Ray}
\email{rajarshi@jcbose.ac.in}
\author{Kinkar Saha}
\email{saha.k.09@gmail.com}
\author{Sudipa Upadhaya}
\email{sudipa.09@gmail.com}
\affiliation{Center for Astroparticle Physics \&
Space Science, Block-EN, Sector-V, Salt Lake, Kolkata-700091, INDIA 
 \\ \& \\ 
Department of Physics, Bose Institute, \\
93/1, A. P. C Road, Kolkata - 700009, INDIA}


\begin{abstract}

The Polyakov$-$Nambu$-$Jona-Lasinio model has been quite successful in
describing various qualitative features of observables for strongly
interacting matter, that are measurable in heavy-ion collision
experiments. The question still remains on the quantitative
uncertainties in the model results. Such an estimation is possible only
by contrasting these results with those obtained from first principles
using the lattice QCD framework.  Recently a variety of lattice QCD data
were reported in the realistic continuum limit. Here we make a first
attempt at reparametrizing the model so as to reproduce these lattice
data. We find excellent quantitative agreement for the equation of
state. Certain discrepancies in the charge and strangeness
susceptibilities as well as baryon-charge correlation still remain. We
discuss their causes and outline possible directions to remove them.

\end{abstract}
\maketitle

\section{Introduction}
\label{sc.intro}

\par
Thermodynamic properties of strongly interacting matter under extreme
conditions is being actively studied theoretically as well as
experimentally. A first principle approach is provided by the finite
temperature formulation of quantum chromodynamics (QCD) on a space-time
lattice. For light quarks these studies
~\cite{Boyd,Engels,Katz,Szabo,Hands,Laermann,Philipsen1,Aoki,Aoki1,
yaoki,Katz1}
indicate the possibility of a rapid crossover between the color confined
and deconfined states. The chiral symmetry is also broken/restored
spontaneously along with the confinement/deconfinement transition.  For
the physical case of two light quarks and a heavy strange quark, lattice
QCD simulations for zero net conserved charges find this cross-over
temperature to be in the range $150~{\rm MeV} < T_c < 160~{\rm MeV}$ as
reported by the Hot-QCD~\cite{Bazavov12,Bazavov14} and
Wuppertal-Budapest~\cite{Borsanyi14} collaborations. A cross-over
transition does not leave a singular boundary between two different
phases. Nevertheless, near $T_c$ various thermodynamic quantities
exhibit a rapid change.  Fluctuations of conserved charges are prominent
quantities in this regard ~\cite{Ejiri,Hatta,Stephanov}. Lattice
simulation results undoubtedly serve as a benchmark estimate over a large
temperature window ~\cite{Karsch1}.

\par
At the same time, it is also important to properly explore the QCD phase
diagram to get a flavor of the physics at varying regimes of temperature
and chemical potential. In fact an exciting question that has puzzled
the community is whether there is any phase transition at non-zero
baryon densities for strongly interacting matter. An interesting
possibility associated with this issue is the existence of a critical
end point somewhere on the phase diagram. Unfortunately in lattice QCD
framework certain technical difficulties arise at the non-zero baryon
chemical potentials. Various intelligent techniques exist to circumvent
these difficulties to some extent ~\cite{Katz,Finitemu1,Philipsen1,
Finitemu2,Laermann,Gavai1, Finitemu3,Allton1,Kaczmarek,Endrodi1,
Endrodi2}.

\par
In this context various QCD inspired models are found to be useful in
describing the aspects of strongly interacting matter at arbitrary
temperature and chemical potentials. In the present article the various
thermodynamic properties of strongly interacting matter are investigated
within the framework of Polyakov loop enhanced Nambu$-$Jona-Lasinio
(PNJL) model. One of the two key ingredients in the PNJL model is the
Nambu$-$Jona-Lasinio (NJL) model
~\cite{YNambu,Hatsuda2,Vogl,Klevansky,Hatsuda1,Buballa1,Barducci}.  This
model includes the global symmetries of QCD in the fermionic sector,
like the chiral symmetry, baryon number, electric charge, strange number
symmetries etc.  The multi-quark interactions in this model are
responsible for the dynamical generation of mass, leading to spontaneous
breaking of chiral symmetry.  However, the gluon fields being integrated
out, this model does not have an adequate mechanism of confinement,
especially for non-zero temperatures. To this end the PNJL
model~\cite{Ogilvie,Fukushima,Ratti} gives a sense of confinement by
introduction of a temporal background gluon field along with its self
interactions mimicking the pure glue effects. Thus by construction both
chiral and deconfinement transitions are entwined within a single
framework.

\par
Interestingly a reasonable parametrization of the PNJL model could be
achieved to obtain qualitatively similar results as in lattice QCD
framework almost a decade
ago~\cite{Ratti,Ray,Mukherjee,Robner,Ghosh,Datta,Fu,Wu}.  Since then
several studies were done to analyze the properties of this model as
well as to improve the model step by step. Improvements in the model for
inclusion of eight-quark interactions in the NJL part
~\cite{Osipov,Hiller,Moreira,Blin} necessary in order to stabilize the
ground state, were introduced in ~\cite{Bhattacharyya,Deb,Lahiri}. In
ref.~\cite{Majumder}, the first case study of the phase diagram in
$\beta$-equilibrium has been reported using the PNJL model.  In a recent
work~\cite{Mustafa} the SU(3) color singlet ensemble of a quark-gluon
gas has been shown to exhibit a Z(3) symmetry and within stationary
point approximation it becomes equivalent to the Polyakov loop ensemble.
In ref.~\cite{Raha} it was shown that though in general a small amount
of mass difference between the two light quarks does not affect the
thermodynamics of the system much, it might have a significant effect on
baryon-isospin correlations.  Studies of various thermodynamic
quantities and fluctuation and correlations of conserved charges
incorporating finite volume effects have been reported in
ref.~\cite{Sur,Bhatta}. Also the first model study of the net charge
fluctuations in terms of D-measure from the PNJL model~\cite{Kinkar} has
been reported.  In an interesting exercise, the validity of the
fluctuation-dissipation theorem has been discussed in the context of the
PNJL model~\cite{Anirban}. As we know, viscous effects play pivotal role
in the evolution of the hot and dense system.  Study of these effects in
terms of transport coefficients have been done in the NJL and PNJL model
~\cite{Redlich,Sabyasachi,Marty,Weise,Shi-Song, Sudipa,Saha,Kaiser} and
compared with hadron resonance gas studies~\cite{Kadam,Mohanty,
Das,Krein}. In ref.~\cite{Claudia,Friman,Fuku,Kahara,Zhang,
Ruivo,Kashiwa,Buballa,Hansen, Lourenco,Abhijit,Inagaki,Friesen}, the
authors have discussed behavioral pattern of different observables as
extracted from the PNJL model. The QCD phase structure has also been
investigated for imaginary chemical potentials in
ref.~\cite{Sakai,Yahiro,Morito} under PNJL model framework.  Different
interesting features of the Polyakov loop have instigated the
development of different formalisms of the PNJL model
~\cite{Salcedo,Sal,Salc,Islam}.  Effects of consideration of gluon
Polyakov loop have been discussed in ref.~\cite{Meg,Megi,Tsai}.
An important set of work being carried out recently are the improvements
of the Polyakov loop potential by introducing the effects of
back-reaction of the quarks, that are supposed to give a more realistic
systematics of full QCD \cite{braun,rincon}. We shall however
restrict ourselves to the simplistic pure glue form of the Polyakov loop
potential with the quark back-reaction essentially coming through the
changes of the mean fields and model parameters.

\par
Given that one of the most important application of the PNJL model would
be to predict observables for non-zero baryon densities, it is important
to at least reproduce observables for zero baryon densities where first
principle results from lattice QCD are available. The qualitative
agreement of results in the PNJL model with those available from lattice
QCD has so far been quite satisfactory. The agreement seemed to be more
convincing once the temperature dependent observables were plotted
against $T/T_c$, where $T_c$ in the model was not equal to that obtained
on the lattice. However the lattice data used for these studies were at
finite lattice spacings. Recently continuum extrapolations for a number
of observables have been reported from lattice simulations. Therefore it
is high time that one tries to set model parameters such as to reproduce
the quantitative agreement of observables with the lattice results. In
the present work we attempt to reset the PNJL model parameters to
reproduce the $T_c$ as well as the temperature dependence of pressure as
obtained in the continuum limit of lattice QCD. Various other
thermodynamic observables may then be obtained from appropriate
derivatives of pressure and contrasted against the lattice QCD results.
The parameters we shall modify are the ones for the Polyakov loop
potential as the parameters of the NJL model are fixed at zero
temperature and densities.

\par
We organize the manuscript as follows. In the next section we describe
the PNJL model focusing on the construction of the effective potential
and the constraints on various parameters.  In section
\ref{sc.parameter} we detail the parameter fixing procedure. Thereafter
we present some thermodynamic quantities in section \ref{sc.thermo} and
discuss the fluctuations and correlations of conserved charges in
section \ref{sc.fluc}. In the final section we summarize and conclude.

\section{PNJL Model}
\label{sc.model}

\par
PNJL model was initialized  with a Polyakov loop effective potential
being added to the NJL model~\cite{Ogilvie,Fukushima,Ratti}. While the
chiral properties are taken care of by the NJL part, the Polyakov loop
explains the deconfinement physics. Extensive studies have been carried
out using PNJL model with 2 and 2+1 flavors
~\cite{Ratti,Ray,Mukherjee,Ghosh,ciminale,Bhattacharyya,Shao,Tang,
Peixoto,Haque}. Here, we consider 2+1 flavor PNJL model taking up to six
and eight quark interaction terms as in ~\cite{ciminale,Bhattacharyya}.
The thermodynamic potential in this case reads as,

\small
\begin{eqnarray}
\Omega (\Phi,\bar{\Phi},\sigma_f,T,\mu ) &=&
2g_S\sum_{f=u,d,s}\sigma_f^2 -
\frac{g_D}{2}\sigma_u\sigma_d\sigma_s + 
3\frac{g_1}{2}(\sum_f\sigma_f^2)^2 + 3g_2\sum_f\sigma_f^4
- 6\sum_f\int_0^\infty\frac{d^3p}{(2\pi)^3}E_f
\Theta(\Lambda-|\vec{p}|)
\nonumber \\
&& - ~2T\sum_f\int_0^\infty\frac{d^3p}{(2\pi)^3} \ln{\left[ 
1+3\left( \Phi+\bar{\Phi} e^{-\left( E_f-\mu_f \right) /T} \right)
e^{-\left( E_f-\mu_f \right) /T } + 
e^{-3\left( E_f-\mu_f \right) /T } \right]  }
\nonumber \\
&& - ~2T\sum_f\int_0^\infty\frac{d^3p}{(2\pi)^3} \ln{\left[ 
1+3\left( \bar{\Phi}+\Phi e^{-\left( E_f+\mu_f \right) /T} \right)
e^{-\left( E_f+\mu_f \right) /T } + 
e^{-3\left( E_f+\mu_f \right) /T } \right]  }
\nonumber \\
&&
+ ~\mathcal{U'}(\Phi,\bar{\Phi},T)
\label{eq.Potential}
\end{eqnarray}
\normalsize

\noindent
The fields $\sigma_f=\langle\bar\psi_f\psi_f\rangle$ correspond to
the two light flavor ($f=u,d$) condensates and the strange ($f=s$) quark
condensate respectively. There is a four quark coupling term with
coefficient $g_S$, a six quark coupling term breaking the axial U(1)
symmetry explicitly with a coefficient $g_D$, and eight quark coupling
terms with coefficients $g_1$ and $g_2$ necessary to sustain a stable
minima in the NJL Lagrangian.  The corresponding quasiparticle energy
for a given flavor $f$ is $E_f=\sqrt{p^2+M_f^2}$, with the dynamically
generated constituent quark masses given by,

\begin{equation}
M_f=m_f-2g_S\sigma_f+\frac{g_D}{2}\sigma_{f+1}\sigma_{f+2}-2g_1\sigma_f
(\sigma_u^2+\sigma_d^2+\sigma_s^2)-4g_2\sigma_f^3
\end{equation} 

\noindent
In the above, if $\sigma_f=\sigma_u$, then $\sigma_{f+1}=\sigma_d$ and
$\sigma_{f+2}=\sigma_s$, and so on in a clockwise manner.  The finite
range integral gives the zero point energy. The different parameters as
obtained from ~\cite{Bhattacharyya} are given in Table~\ref{tb.njlpara}.

\begin{table}[!htb]
\begin{tabular}{|c|c|c|c|c|c|c|c|}
\hline
\hline
Interaction & $m_u$ (MeV) & $m_s$ (MeV) & $\Lambda$ (MeV) &
$g_s\Lambda^2$ & $g_D\Lambda^5$ & $g_1\times 10^{-21}$ (MeV$^{-8}$) &
$g_2\times 10^{-22}$ (MeV$^{-8}$) \\
\hline
\hline
6-quark &  5.5 & 134.758 & 631.357 & 3.664 & 74.636 & 0.0  & 0.0 \\
\hline
8-quark & 5.5  & 183.468 & 637.720 & 2.914 & 75.968 & 2.193 & -5.890 \\
\hline
\end{tabular}
\caption{Parameters in the NJL model}
\label{tb.njlpara}
\end{table}

\noindent
The finite temperature and chemical potential contributions of the
constituent quarks are given by the next two terms. Note that these are
basically coming from the fermion determinant in the NJL model modified
due to the presence of the fields corresponding to the traces of
Polyakov loop and its conjugate given by $\Phi=\frac{Tr_cL}{N_c}$ and
$\bar{\Phi}=\frac{Tr_cL^{\dagger}}{N_c}$ respectively. Here
$L(\vec{x})=\mathcal{P} exp\left[ i\int_0^{1/T}d\tau A_4(\vec{x},\tau)
\right]$ is the Polyakov loop, and $A_4$ is the temporal component
of background gluon field.

\par
The effective potential that describes the self interaction of the
$\Phi$ and $\bar{\Phi}$ fields are given by $\mathcal{U'}$. Various
forms of the potential exist in the literature (see e.g.
~\cite{Robner,Fuku,Ghosh,Contrera,Qin}). We shall use the form
prescribed in~\cite{Ghosh} which reads as,

\begin{equation}
\frac{\mathcal{U'}(\Phi,\bar{\Phi},T)}{T^4}=
\frac{\mathcal{U}(\Phi,\bar{\Phi},T)}{T^4}-\kappa
ln[J(\Phi,\bar{\Phi})].
\label{eq.Ppotential}
\end{equation}

\noindent
Here ${\mathcal U}(\Phi,\bar{\Phi},T)$ is a Landau-Ginzburg type
potential commensurate with the global Z(3) symmetry of the Polyakov
loop~\cite{Ratti}. $J(\Phi,\bar{\Phi})$ is the Jacobian of transformation
from the Polyakov loop to its traces, and $\kappa$ is a dimensionless
parameter which is determined phenomenologically.  The effective
potential is chosen to be of the form,

\begin{equation}
\frac{\mathcal{U}(\Phi,\bar{\Phi},T)}{T^4}=
-\frac{b_2(T)}{2}\bar{\Phi}\Phi-\frac{b_3}{6}
(\Phi^3+\bar{\Phi}^3)+\frac{b_4}{4}(\bar{\Phi}\Phi)^2
\end{equation}

\noindent
The coefficient $b_2(T)$ is chosen to have a temperature dependence of
the form,

\begin{equation}
b_2(T)=a_0+a_1exp(-a_2\frac{T}{T_0})\frac{T_0}{T},
\end{equation}

\noindent
and $b_3$ and $b_4$ are chosen to be constants. In the next section we
discuss the methodology for fixing these parameters.

\section{Fixing Polyakov loop potential parameters}
\label{sc.parameter}

\par
The Polyakov loop fields are expected to approach unity for large
temperatures. Therefore for an effective model of pure glue theory 
the minimization of the potential would be obtained for 
$lim_{T\rightarrow\infty}\Phi=1$. Also the pressure should be that of
the massless free gluon gas. Using these two conditions one may obtain
$b_3$ and $b_4$ in terms of $b_2(T\rightarrow\infty)=a_0$. The
parameters $a_1$, $a_2$ and $T_0$ and $\kappa$ may thereafter be fixed
phenomenologically by requiring that the crossover temperature comes
around $T_c\sim 160 {\rm MeV}$, along with the pressure to agree with the
lattice QCD results for various temperatures.

\begin{table}[!htb]
\begin{tabular}{|c|c|c|c|c|c|c|c|}
\hline \hline
Interaction & $T_0$ (MeV) & $a_0$ & $a_1$ & $a_2$ & $b_3$ & $b_4$ &
$\kappa$\\
\hline
6-quark & 175 & 6.75 & -9.0 & 0.25 & 0.805 & 7.555 & 0.1 \\
\hline
8-quark & 175 & 6.75 & -9.8 & 0.26 & 0.805 & 7.555 & 0.1    \\
\hline \hline
\end{tabular}
\caption{Parameters for the Polyakov loop potential.}
\label{tb.polpara}
\end{table}

\par
We first fixed the parameter values of $a_0$, $T_0$ and $\kappa$. Then $b_3$
and $b_4$ were obtained in terms of $a_0$. Thereafter $a_1$ and $a_2$
were adjusted to get the best combination for the crossover temperature
$T_c$ and the pressure vs temperature plot to agree with continuum limit
obtained from lattice QCD computations.  The set of parameters thus
obtained is given in Table~\ref{tb.polpara}.

\begin{figure}[!htb]
{\includegraphics[height=8cm,width=6.0cm,angle=270]
{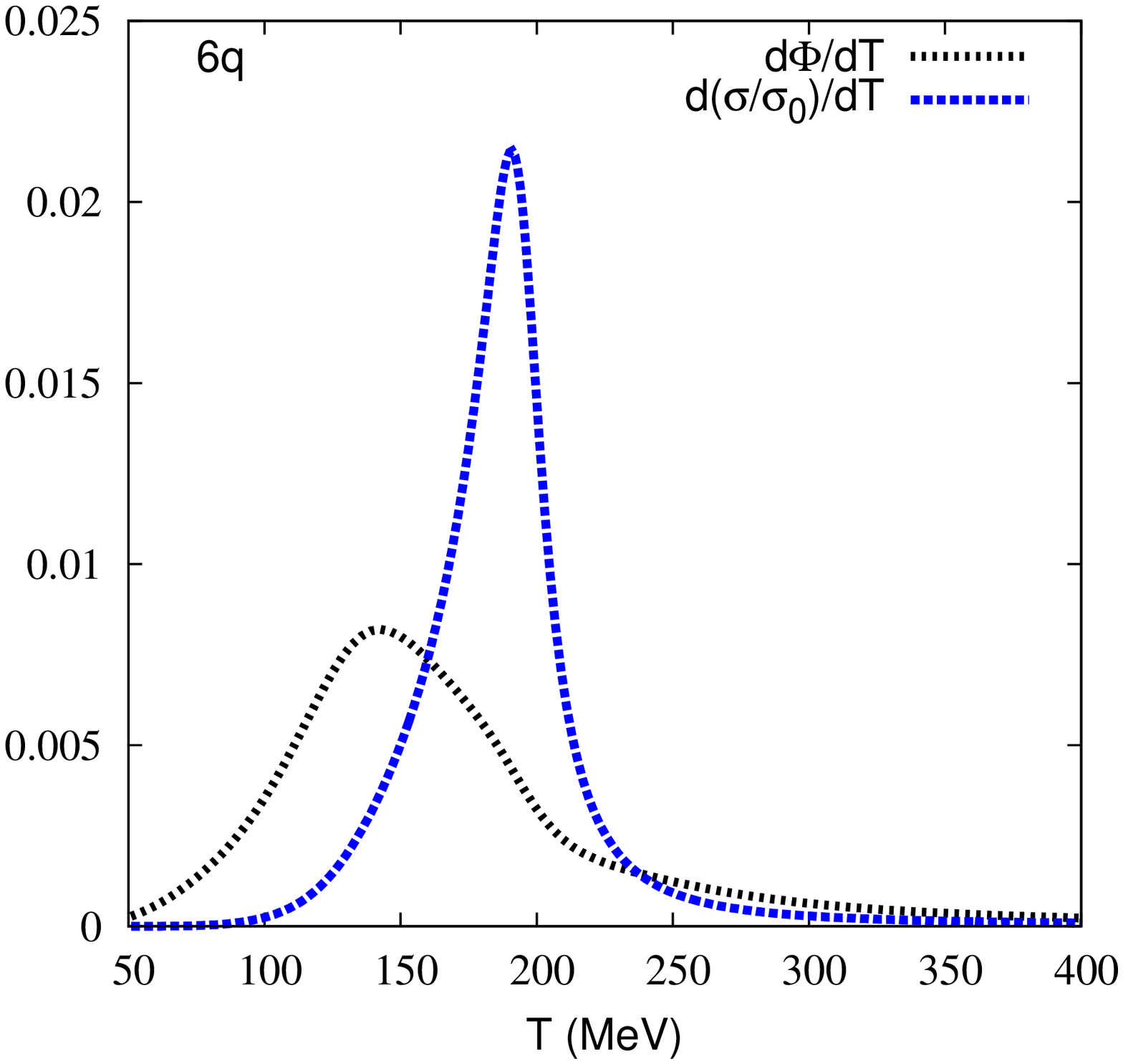}}
{\includegraphics[height=8cm,width=6.0cm,angle=270]
{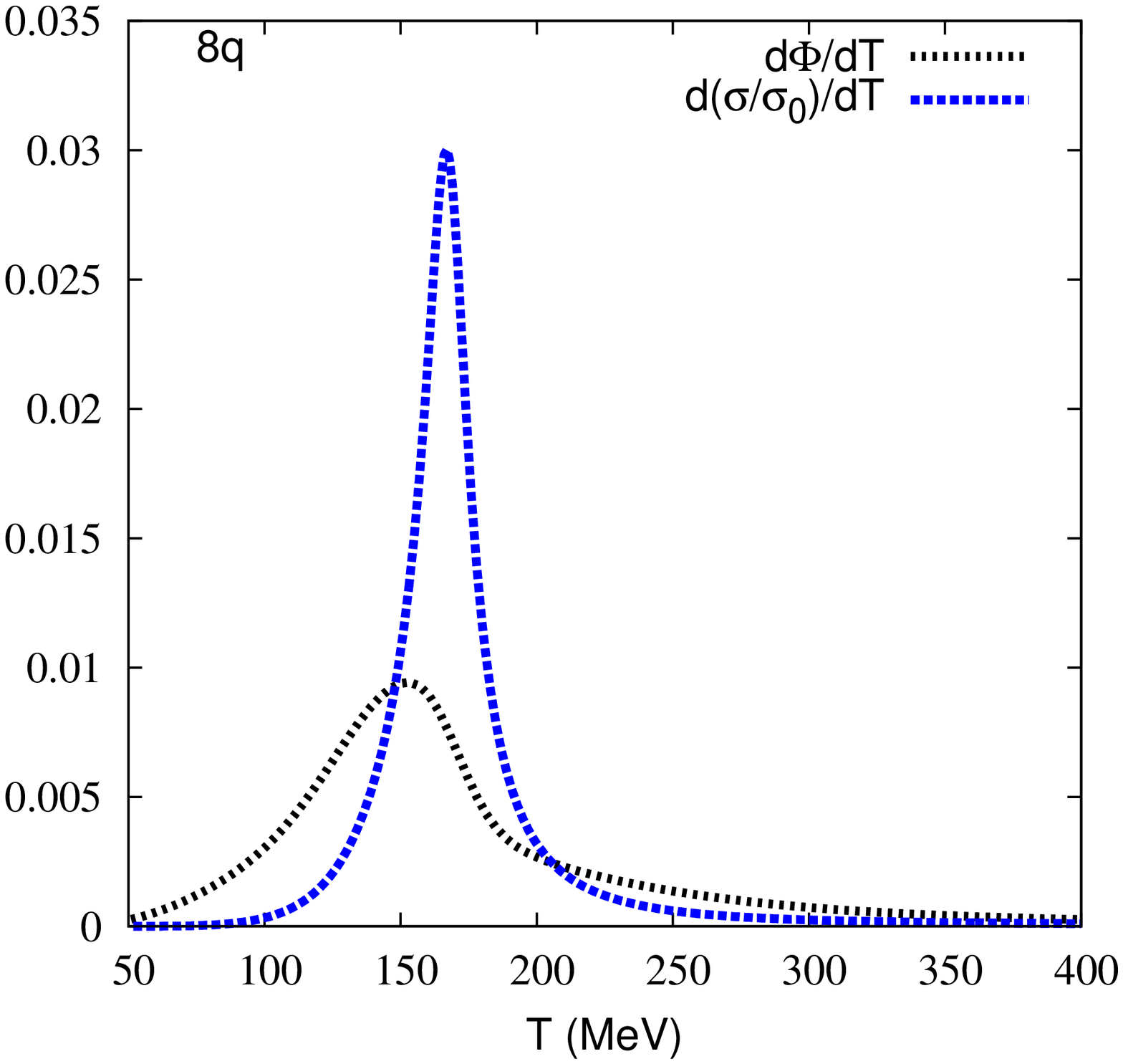}}
\caption{(color online) The temperature derivatives of $\sigma_f$ for
light flavors and Polyakov loop fields for 6-quark (left) and 8-quark
(right) interactions. $\sigma_0$ denotes the value of $\sigma_f$ at
$T=0$.} 
\label{fg.fderiv}
\end{figure}

\par
The deconfinement temperature obtained in lattice QCD with physical
quark masses from the fluctuation of the Polyakov loop is much higher
than the chiral transition temperature \cite{Aoki,yaoki}. However the
deconfinement temperature as measured from the peak of the entropy of a
static quark is found to be consistent with the chiral transition
temperature \cite{jweber}. In our model framework we consider the
temperature derivatives of the mean fields and locate their peaks to
obtain the transition temperature. The temperature derivative of the
Polyakov loop is closely related to the definition of temperature
derivative of the static quark free energy that gives its entropy as
defined in \cite{jweber}.  The plots for $d\sigma_f/dT$ for light
flavors and $d\Phi/dT$ are shown in Fig.~\ref{fg.fderiv}. The
corresponding $T_c$ was obtained from the average of the two peak
positions. It was observed that the modification of parameter values did
not produce any appreciable reduction of the $T_c$ from what we have
obtained. This means that with only the adjustments of parameters of the
Polyakov loop potential $T_c$ cannot be reduced further. In fact there
is a drastic reduction in $T_c$ for 6-quark interactions here compared
to our earlier parametrization reported in~\cite{Bhattacharyya}. The
reduction is quite small for the 8-quark interaction. The resulting
values of $T_c$ are listed in Table \ref{tb.tc}.

\begin{table}[!htb]
\begin{tabular}{|c|c|c|c|}
\hline \hline
Interaction & Peak position of $d\Phi/dT$ (MeV) &
Peak position of $d\sigma/dT$ (MeV) & $T_c$ (MeV) \\
\hline
6-quark & 142 & 191 & 166.5 \\
\hline
8-quark & 158 & 167 & 162.5 \\
\hline \hline
\end{tabular}
\caption{Location of crossover temperature}
\label{tb.tc}
\end{table}

\begin{figure}[!htb]
{\includegraphics[height=8cm,width=6.0cm,angle=270]
{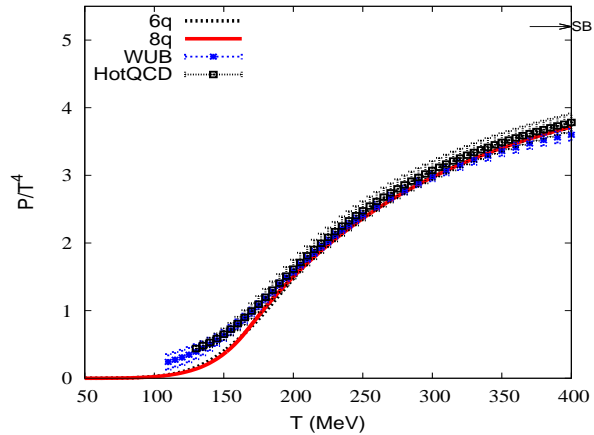}}
\caption{(color online) Variation of pressure scaled with $T^4$ as
function of temperature.  The continuum extrapolated dataset of HotQCD
~\cite{Bazavov14} and Wuppertal-Budapest (WUB)~\cite{Borsanyi14}
collaborations are shown.}
\label{fg.pressbyT4}
\end{figure}

\par
In Fig.~\ref{fg.pressbyT4} we show scaled pressure, as a function of
temperature. The scaled pressure grows from close to zero at small
temperatures and reaches almost 75$\%$ of the Stefan-Boltzmann (SB)
limit commensurate with present day continuum lattice data
~\cite{Bazavov14,Borsanyi14}. This is in sharp contrast to the earlier
results in which the scaled pressure was shown to grow to almost 90$\%$
of the SB limit~\cite{Bhattacharyya}, commensurate with finite lattice
spacing data available at that time~\cite{Cheng08}. Thus by refixing the
parameters of the Polyakov loop potential we have been able to achieve
both a crossover temperature of $T_c\sim160~{\rm MeV}$ as well as
quantitative agreement of temperature variation of pressure with the
lattice QCD continuum estimation.  For temperatures below $T_c$ the
model results do differ slightly from the lattice data.  We note that
though the lattice data by the Hot-QCD and Wuppertal-Budapest group
agree within error bars for the lower values of temperature there is
about a standard deviation of difference for the higher temperature
ranges. We simply adjusted the parameters so that in the PNJL model the
pressure goes through values from one of them chosen randomly - in this
case the Hot-QCD data. We also note that there is almost no difference
between pressure vs temperature plot of the 6-quark and 8-quark
interaction versions of the PNJL model by construction.

\section{THERMODYNAMICS}
\label{sc.thermo}

\begin{figure}[!htb]
{\includegraphics[height=8cm,width=6.0cm,angle=270]
{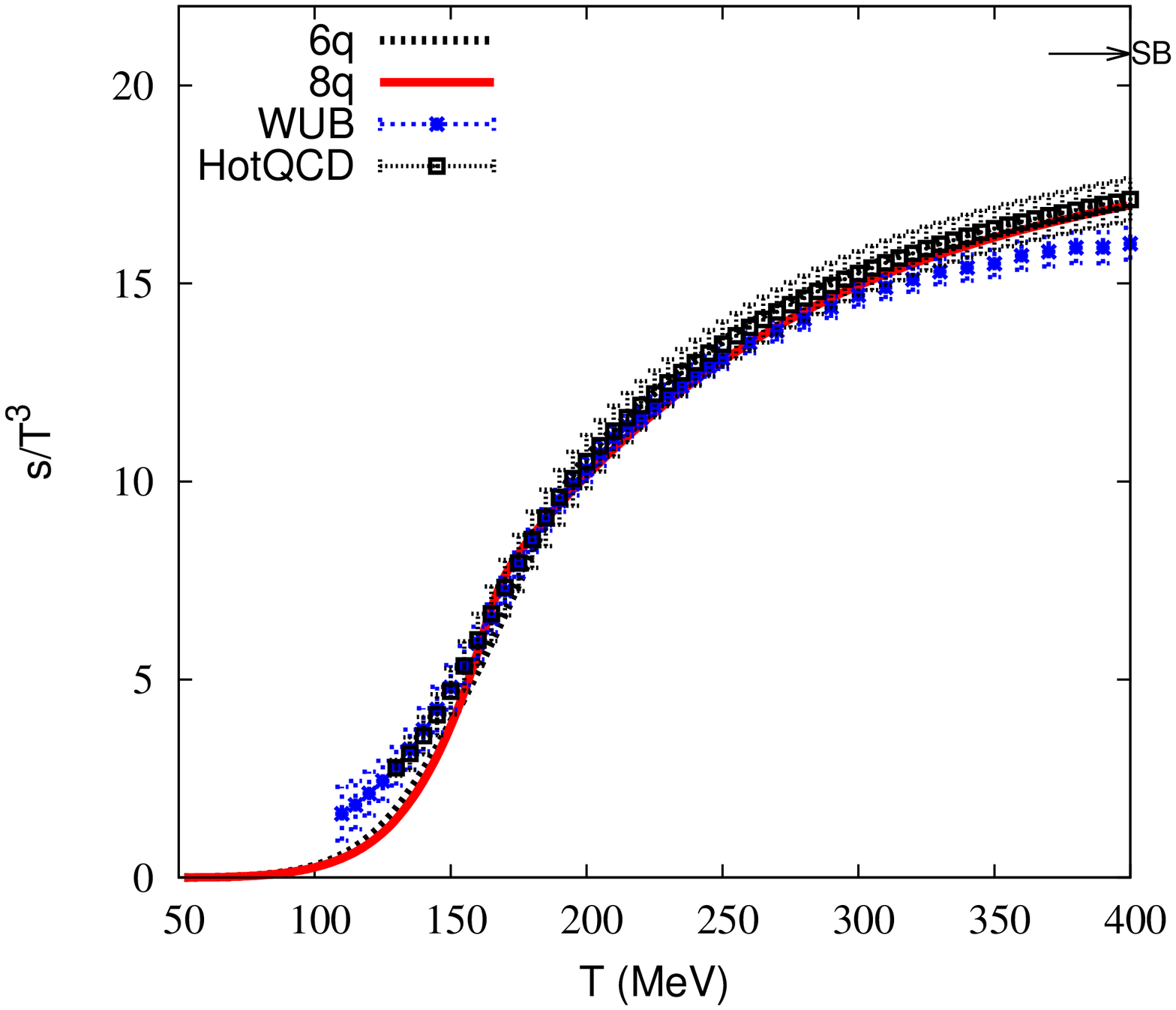}}
{\includegraphics[height=8cm,width=6.0cm,angle=270]
{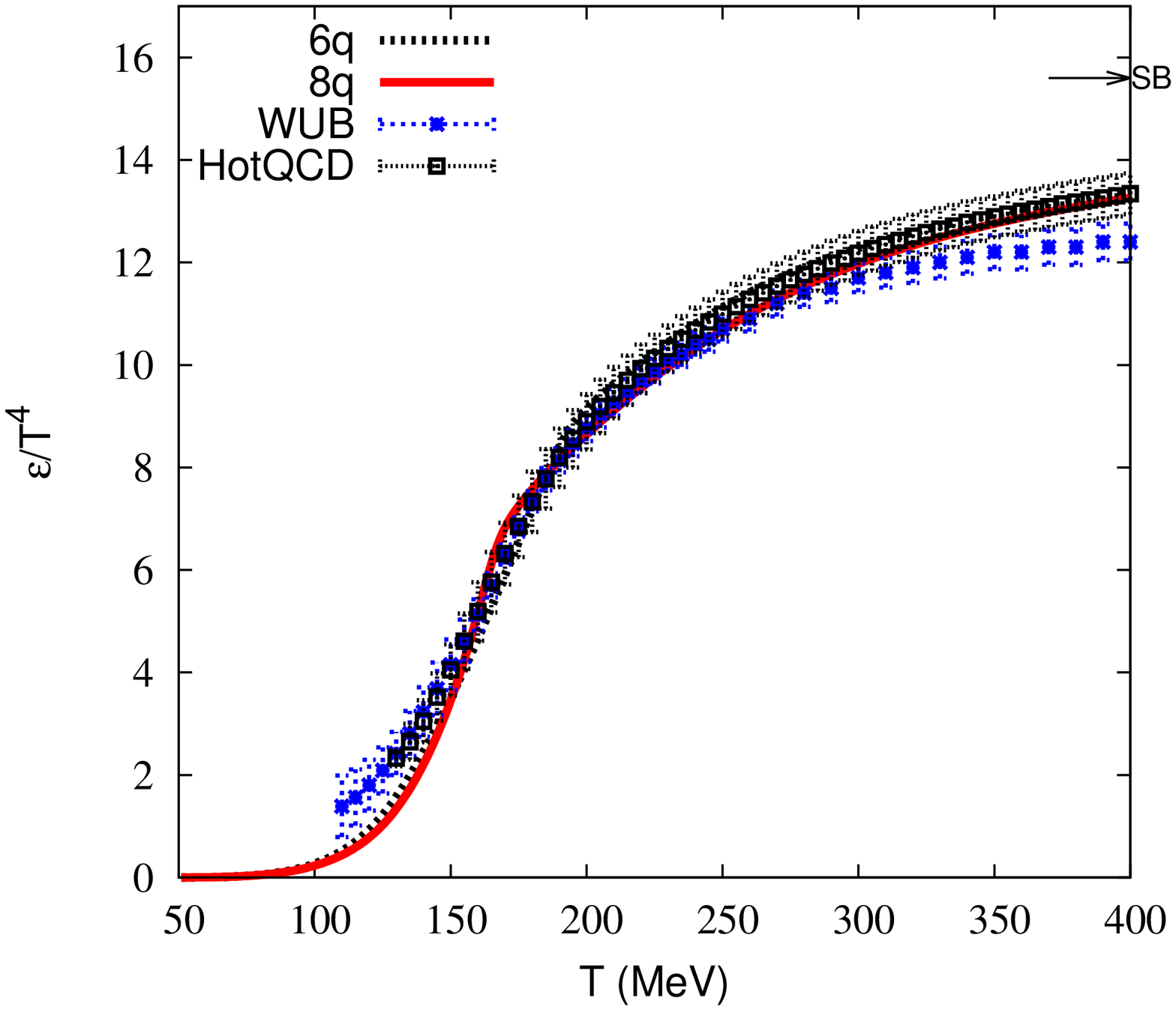}}
\caption{(color online) The scaled entropy (left) and scaled energy
density (right) as functions of temperature.  The continuum extrapolated
dataset of HotQCD and Wuppertal-Budapest (WUB) collaborations are taken
respectively from ~\cite{Bazavov14} and ~\cite{Borsanyi14}.}
\label{fg.enteng}
\end{figure}

\par
The various thermodynamic quantities can now be obtained from
corresponding derivatives of pressure that arise from the respective
thermodynamic relations. From the first order derivative of pressure
with respect to temperature, one can obtain the entropy density
$s=\frac{\partial P}{\partial T}$ and energy density
$\epsilon=T^2\frac{\partial (P/T)}{\partial T} =T\frac{\partial
P}{\partial T}-P$. These are plotted in Fig.~\ref{fg.enteng}. They are
also contrasted with recent Hot-QCD and Wuppertal-Budapest continuum
results. We find that the results of PNJL model satisfactorily reproduce
lattice data quantitatively.  Here again the difference between the two
sets of lattice QCD data at high temperatures are evident, and our
results align well with the Hot-QCD data by construction. For $T<T_c$
the PNJL results deviate from lattice QCD data by a small amount similar
to that observed for pressure.

\begin{figure}[!htb]
{\includegraphics[height=8cm,width=6.0cm,angle=270]
{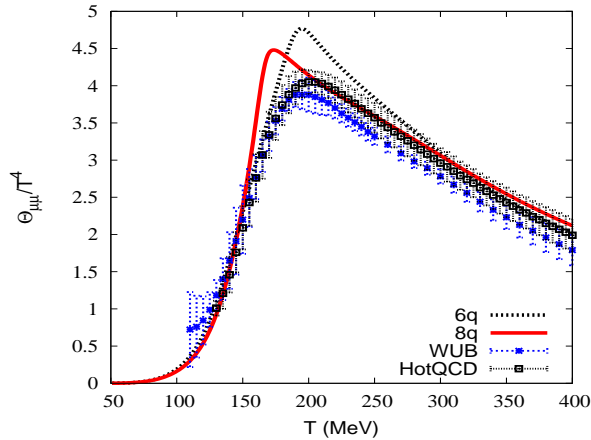}}
\caption{(color online) Trace of energy momentum tensor as a function of
temperature. The continuum extrapolated dataset of HotQCD and
Wuppertal-Budapest (WUB) collaborations are taken respectively from
~\cite{Bazavov14} and ~\cite{Borsanyi14}.}
\label{fg.trace}
\end{figure}

\par
Given that the equation of state in the PNJL model agrees well with the
lattice data we now consider other observables. The energy-momentum
tensor $\Theta_{\mu\mu}=\epsilon-3P$ obtained in the PNJL model has a
small difference with the lattice data near $T_c$ as shown in
Fig.~\ref{fg.trace}.  In fact there is a similar small difference
between the 6-quark and 8-quark versions of the PNJL model. But the
overall agreement over the full range of temperature is quite
satisfactory. Comparing to earlier estimates based on finite lattice
spacings it may be noted that the quantitative value of the height of
the peak here has reduced to almost half of what was reported
in~\cite{Bhattacharyya}.

\begin{figure}[!htb]
{\includegraphics[height=8cm,width=6.0cm,angle=270]
{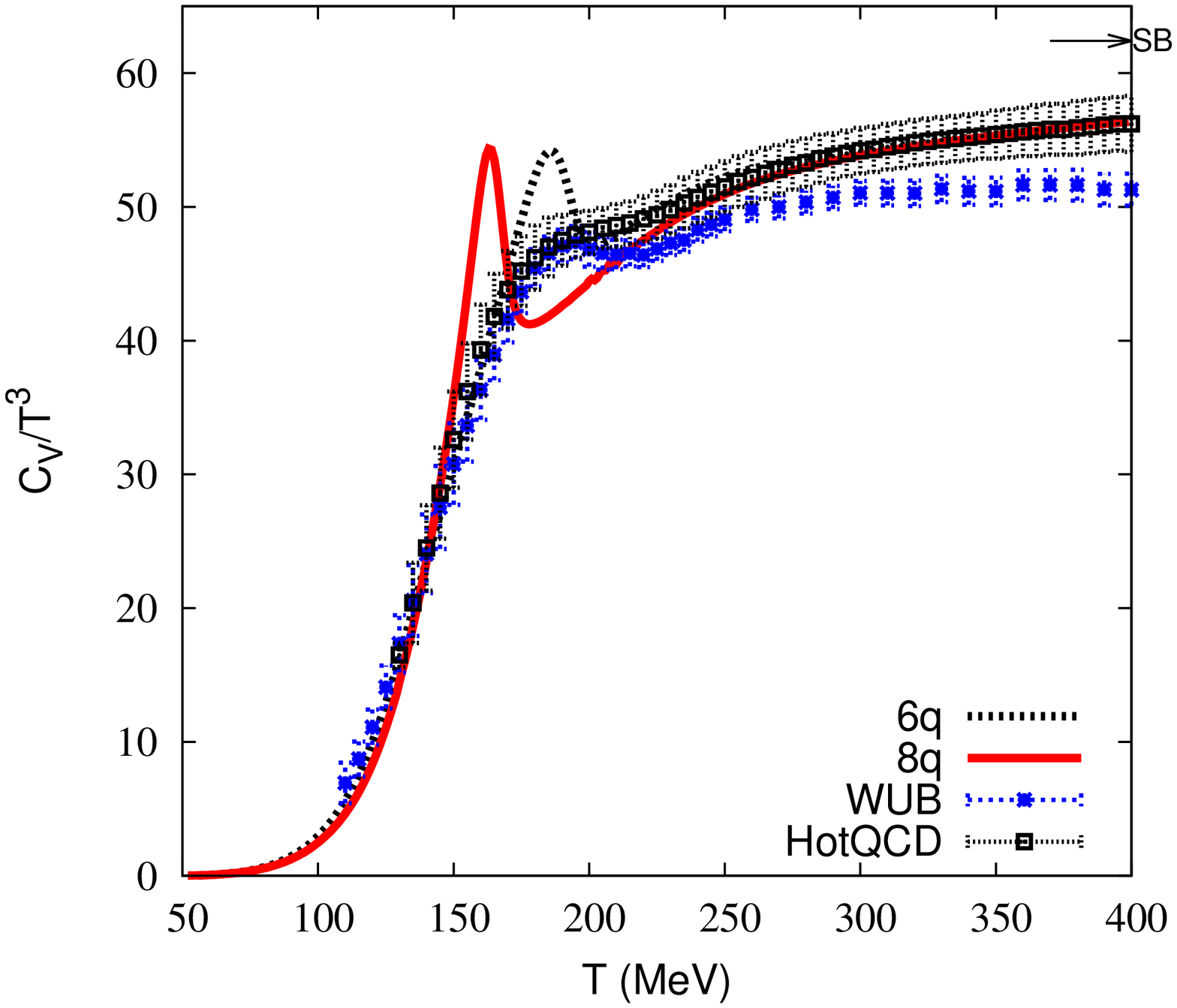}}
{\includegraphics[height=8cm,width=6.0cm,angle=270]
{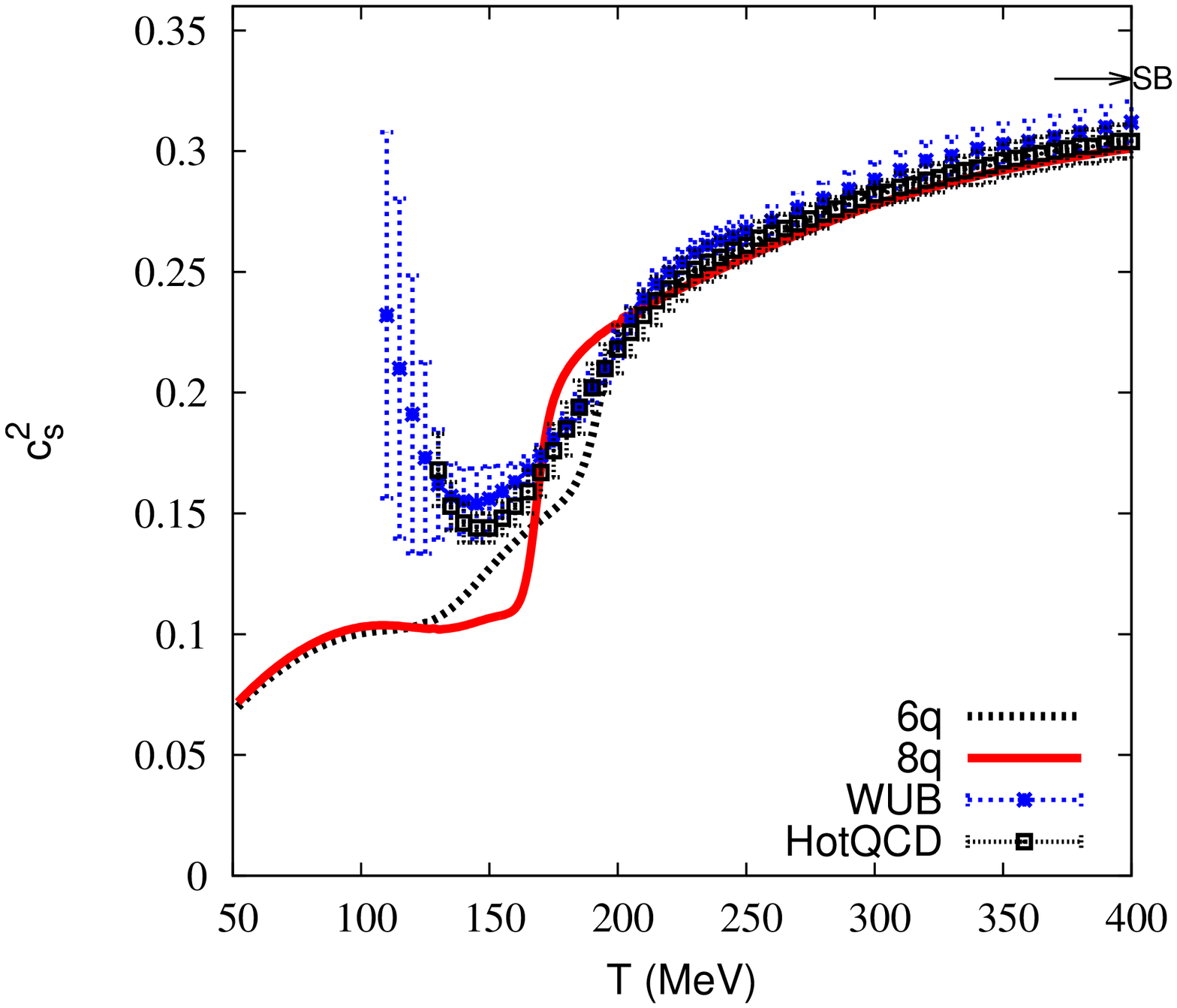}}
\caption{(color online) The specific heat (left) and squared speed of
sound (right) as functions of temperature.  The continuum extrapolated
dataset of HotQCD and Wuppertal-Budapest (WUB) collaborations are taken
respectively from ~\cite{Bazavov14} and ~\cite{Borsanyi14}.}

\label{fg.cvcs}
\end{figure}

\par
From the second order derivative of pressure with respect to temperature
we obtain two important quantities namely the specific heat at constant
volume
$C_V=\frac{\partial \epsilon}{\partial T} =T\frac{\partial ^2P}{\partial
T^2}$, and the squared speed of sound $c_s^2=\frac{\partial P}{\partial
\epsilon} =\frac{s}{C_V}$. These are shown in Fig.~\ref{fg.cvcs}. We
find the specific heat obtained in PNJL model to agree well with the
lattice QCD results except near the crossover region. In this region,
$C_V/T^3$ obtained from the PNJL
model shows a small peak, but the lattice results are completely smooth.
Though the lattice results do not show any peak there is a definite
indication of a hump near the critical region. The differences between
$\Theta_{\mu\mu}$ and $C_V$ obtained in the PNJL model and those on the
lattice indicate that the crossover in the model is somewhat sharper
than that on the lattice. However the size of the peak obtained here is
substantially reduced compared to what was obtained with the earlier
parametrizations~\cite{Deb}, and remains below the SB limit.. 

\par
The temperature variation of the speed of sound is shown in
Fig.~\ref{fg.cvcs}. One expects that at very low temperatures the speed
of sound would be small as the pressure of the system is negligible and
hadrons are massive. With increase in temperature the speed of sound
will increase. However with increasing temperature the hadron resonances
with higher and higher masses would be excited and the speed of sound
would not reach the SB limit. In fact it may even start decreasing with
temperature \cite{Andronic12}. After the crossover the degrees of
freedom change from hadronic to partonic and therefore speed of sound
may again increase.  The minimum of the speed of sound known as the
softest point may be a crucial indicator of the transition to be
observed in heavy-ion collisions \cite{Hung95}.  Such a minimum in the
temperature variation of speed of sound is visible in the lattice QCD
data as shown in Fig.~\ref{fg.cvcs}, but is clearly absent in the PNJL
model results. We note that the PNJL model results are consistent with
the lattice data above $T_c$. The disagreement ensues in the phase where
hadronic degrees of freedom are dominant. The PNJL model in the present
form do not encapsulate the hadronic excitations effectively which has
resulted in this discrepancy. We shall address proper extensions of the
model elsewhere. 

\begin{figure}[!htb]
{\includegraphics[height=8.0cm,width=6.0cm,angle=270]
{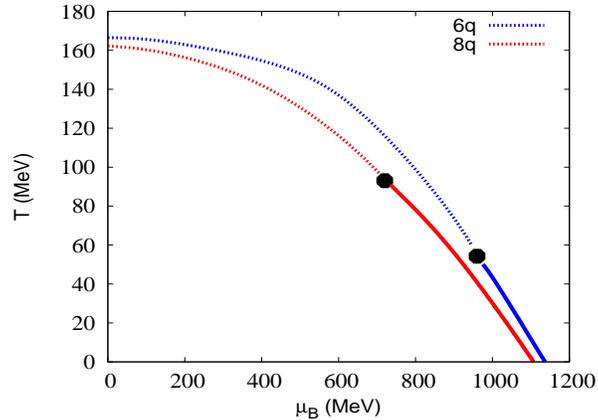}}
\caption{(color online) Phase diagram for 2+1 flavor PNJL with 6 and 8
quark interactions}. 
\label{fg.phasediagram}
\end{figure}

\noindent
$The~Phase~Diagram~:$ 

\par
Exploration of the phase diagram of strongly interacting matter is one
of the major goals of the heavy-ion collision experiments. The currently
running Beam Energy Scan experiments at the RHIC facility
~\cite{starbes}, and the upcoming Compressed Baryonic Matter experiment
at the FAIR facility ~\cite{cbmbook} and the experiments at the NICA
facility \cite{nica} are specifically designed for this purpose.

\par
The phase diagram in the $T-\mu_B$ plane for strongly interacting matter
is being investigated theoretically for quite some time ~\cite{criep}.
While there is a crossover of hadronic phase to partonic phase along the
$T$ direction as suggested by lattice QCD studies, the transition along
the $\mu_B$ direction is expected to be of first order from the various
effective model analysis. The first order line is expected to bend
towards the $T$ axis starting from some finite $\mu_B$ and end at a
critical end point (CEP). This will have some value of temperature $T_E$
and chemical potential ${\mu_B}_E$.

\par
A direct location of the CEP in lattice QCD is spoilt due the appearance
of complex weight factors for non-zero $\mu_B$ in the Monte Carlo
simulations. Several techniques exist that can circumvent this
difficulty to a limited extent. Using a reweighing technique the
location of CEP was estimated first in ~\cite{Fodor1}. Calculations in
the imaginary chemical potential shows conflicting results of existence
of CEP depending on the version of lattice fermions chosen
~\cite{Forcrand1,Jin1}. Radius of convergence analysis for the Taylor
series expansion of pressure may also lead to an estimate of the CEP
~\cite{Hands,Gavai1,Allton1,Ejiri2,Gavai2,Karthik}. However a conclusive
estimate of the CEP does not seem to have been reached. The present
spread in the location of CEP is in the range $0.95T_c<T_E<0.99T_c$ and
$1.5T_c<{\mu_B}_E<2.5T_c$.

\begin{table}[!htb]
\begin{tabular}{|c|c|c|c|c|}
\hline \hline
Interaction & $T_E$ (MeV) &  $T_E/T_c$ & ${\mu_B}_E$ (MeV)
& ${\mu_B}_E/T_c$  \\
\hline
6-quark & 54.3 & 0.326 & 960 & 5.77 \\
\hline
8-quark & 93.0 & 0.572 & 720 & 4.43 \\
\hline \hline
\end{tabular}
\caption{Location of critical end point}
\label{tb.cep}
\end{table}

\par
We have plotted the possible phase diagram in the PNJL model in
Fig.~\ref{fg.phasediagram} considering both 6-quark and 8-quark
interactions. The parameter values are held at those obtained along the
temperature axis. We have used the inflection points i.e. the
temperature derivative of the chiral condensate as well as that of the
Polyakov loop and considered their average as the estimate of the
transition temperature for a given chemical potential. For a first order
transition however we located the point of discontinuous jump of the
field values themselves. At the critical end point the discontinuity
vanishes and the derivative is sharply diverging.

\par
The location of the critical end point is presented in Table
~\ref{tb.cep}. The values are expectedly quite different from those
obtained by us earlier with different set of parameter values
~\cite{Bhattacharyya}. Given that the $T_c$ itself has been decreased by
more than 25 MeV here for the 6-quark interaction, the $T_E$ has reduced
by about 40 MeV. For the 8-quark interaction the $T_c$ value is reduced
here by about 6 MeV, which has resulted in reducing the corresponding
$T_E$ by about 25 MeV. The ${\mu_B}_E$ values are quite large
and differ within 30 MeV for both the interaction models.  The estimates
of the location of CEP obtained from the lattice QCD simulations with
various limitations as summarized in ~\cite{Karsch1}, are still
significantly different from our model estimates.

\section{Fluctuations of conserved charges}
\label{sc.fluc}

\begin{figure}[!htb]
{
\includegraphics[scale=0.38,angle=270]{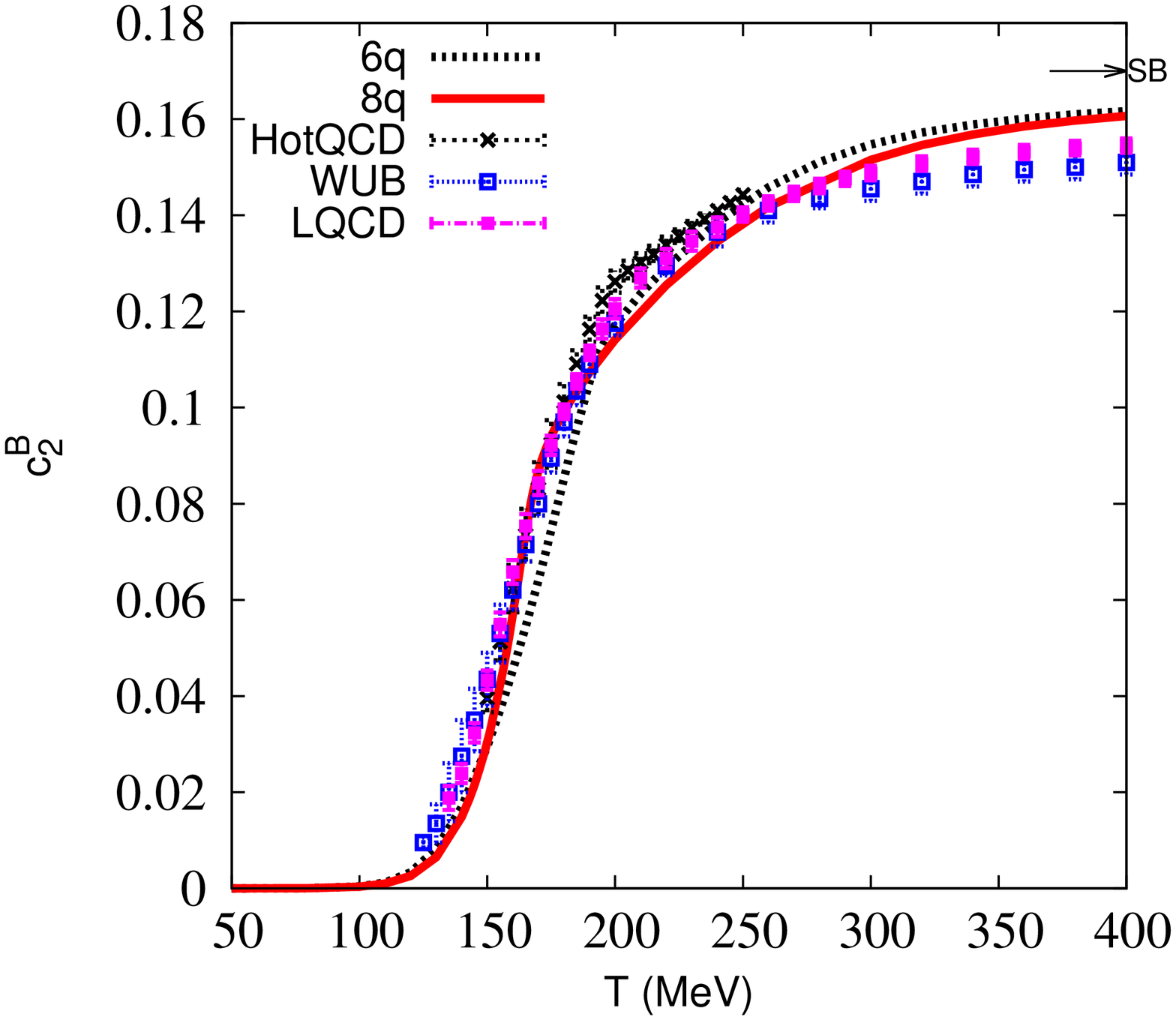}
\includegraphics[scale=0.38,angle=270]{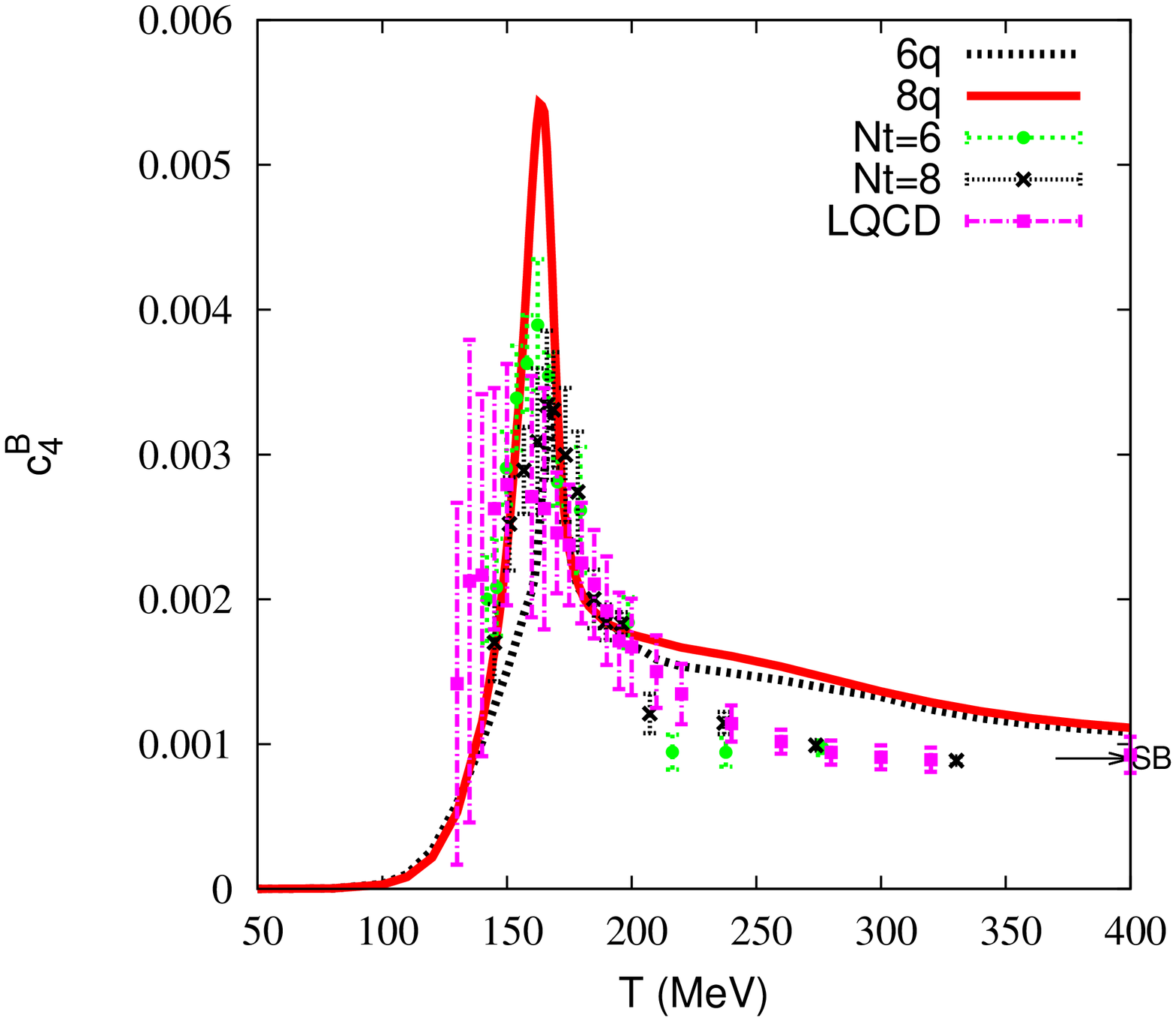}
}
\caption{(color online) Variation of $c_2^B$ and $c_4^B$ as functions of
temperature. The continuum extrapolated dataset for $c_2^B$ of HotQCD
and Wuppertal-Budapest (WUB) collaborations are taken respectively from
~\cite{BazavovPRD} and ~\cite{BorsanyiJHEP}, as well as the data from
~\cite{Bellwied2} denoted as LQCD. For $c_4^B$ HotQCD data for Nt=6 and 8
~\cite{BazavovPRL} and the continuum data from ~\cite{Bellwied2} are
considered.}
\label{fg.cB}
\end{figure}

\par
Fluctuations and correlations of conserved charges are considered
important for their role in determining the state of strongly
interacting matter at high temperatures and densities
~\cite{Koch,Deb,Lahiri,Abelev,Ejiri}. They may also be useful as
signatures of a possible phase transition or crossover
~\cite{Jeon,Jeon1,Heinz,Kinkar,Subhasis,Paramita,
Rajarshi,Subrata,Upadhaya,Hatta,Shuryak}. The pressure of the system at
a given temperature and arbitrary chemical potentials may be expanded as
a Taylor series around zero chemical potentials. The coefficients of
this series are directly related via fluctuation dissipation
theorem~\cite{Anirban}, to the fluctuations and correlations at various
orders.  The basic globally conserved quantities in the strong
interactions are the various flavors considered. These are related to
the experimentally observed charges of baryon number $B$, electric
charge $Q$ and strangeness $S$.  The diagonal Taylor coefficients
$c_n^X(T)$ ($X=B,Q,S$) of $n^{th}$ order in an expansion of the scaled
pressure $P(T,\mu_B,\mu_Q,\mu_S)/T^4$ may be written in terms of the
fluctuations $\chi_n^X(T)$ of the corresponding order as,

\begin{equation}
c_n^X(T)=\frac{1}{n!}\frac{\partial^n(P/T^4)}
{\partial(\frac{\mu_X}{T})^n}=T^{n-4}\chi_n^X(T)
\end{equation}

\noindent
where the expansion is carried out around $\mu_B=0=\mu_Q=\mu_S$. The
off-diagonal coefficients $c_{n,m}^{X,Y}(T)$ ($X,Y=B,Q,S$; $X\ne Y$) in
the $(m+n)^{th}$ order in the Taylor expansion are related to the
correlations between the conserved charges $\chi_{n,m}^{X,Y}(T)$ as,

\begin{equation}
c_{m,n}^{X,Y}=\frac{1}{m!n!}\frac{\partial^{m+n}(P/T^4)}
{(\partial(\frac{\mu_X}{T})^m)(\partial(\frac{\mu_Y}{T})^n)}
=T^{m+n-4}\chi_{n,m}^{X,Y}(T)
\end{equation}

\par
Various fluctuations and correlations of the conserved charges have been
measured in the lattice QCD framework either in the continuum limit
~\cite{BorsanyiJHEP,BazavovPRD,Katz1,Bellwied1,Swagato1,Swagato2,
Bellwied2} or for small lattice spacings, which are expected to be not
far from the continuum limit ~\cite{BazavovPRL}.  Here we present a
comparative study of these quantities with the present parametrization
of the PNJL model. The quantities were obtained in the model by a
suitable Taylor series fitting as has been discussed in detail
in~\cite{Ray}. 

\par
In Fig.~\ref{fg.cB} the variation of the baryon number
susceptibilities $c_2^B$ and $c_4^B$ are shown as functions of
temperature. While $c_2^B$ mimics the behavior of an order parameter,
$c_4^B$ acts as its fluctuation. Apart from the qualitative similarity
with the lattice QCD data, the quantitative agreement is encouraging.
The second order susceptibility $c_2^B$ seems to be impressively close
to the lattice data except for a small difference beyond $T\sim 300 {\rm
MeV}$. Also the difference between the results for the 6-quark and the
8-quark interactions are quite small. For the fourth order
susceptibility $c_4^B$ similar difference remains between the lattice
and model results at the higher temperature region. 

\begin{figure}[!htb]
{
\includegraphics[scale=0.38,angle=270]{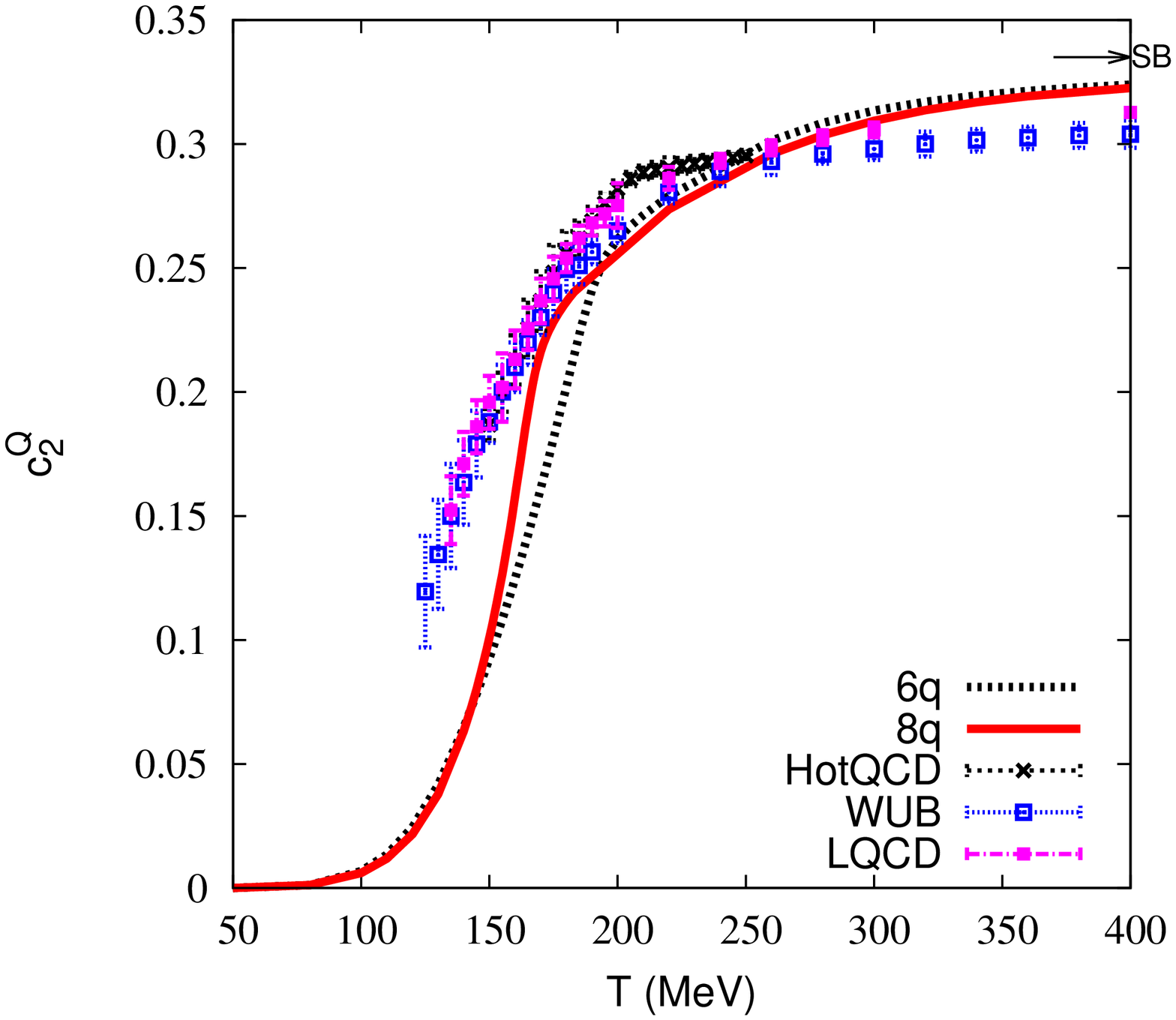}
\includegraphics[scale=0.38,angle=270]{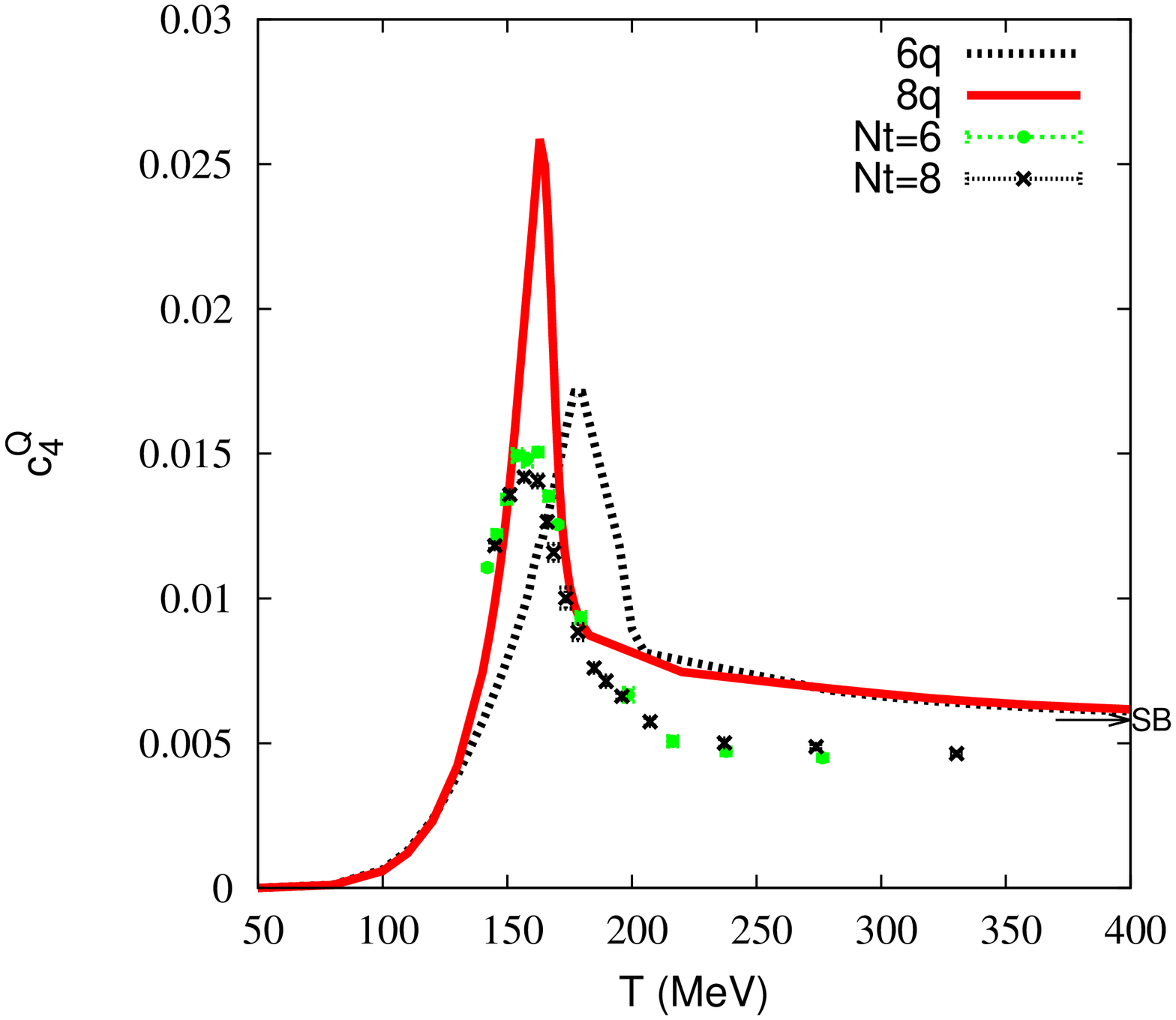}
}
\caption{(color online) Variation of $c_2^Q$ and $c_4^Q$ as functions of
temperature. The continuum extrapolated dataset for $c_2^Q$ of HotQCD
and Wuppertal-Budapest (WUB) collaborations are taken respectively from
~\cite{BazavovPRD} and ~\cite{BorsanyiJHEP}, as well as the data from
~\cite{Bellwied2} denoted as LQCD. For $c_4^Q$ HotQCD data for Nt=6 and 8
~\cite{BazavovPRL}.}
\label{fg.cQ}
\end{figure}

\begin{figure}[!htb]
{
\includegraphics[scale=0.38,angle=270]{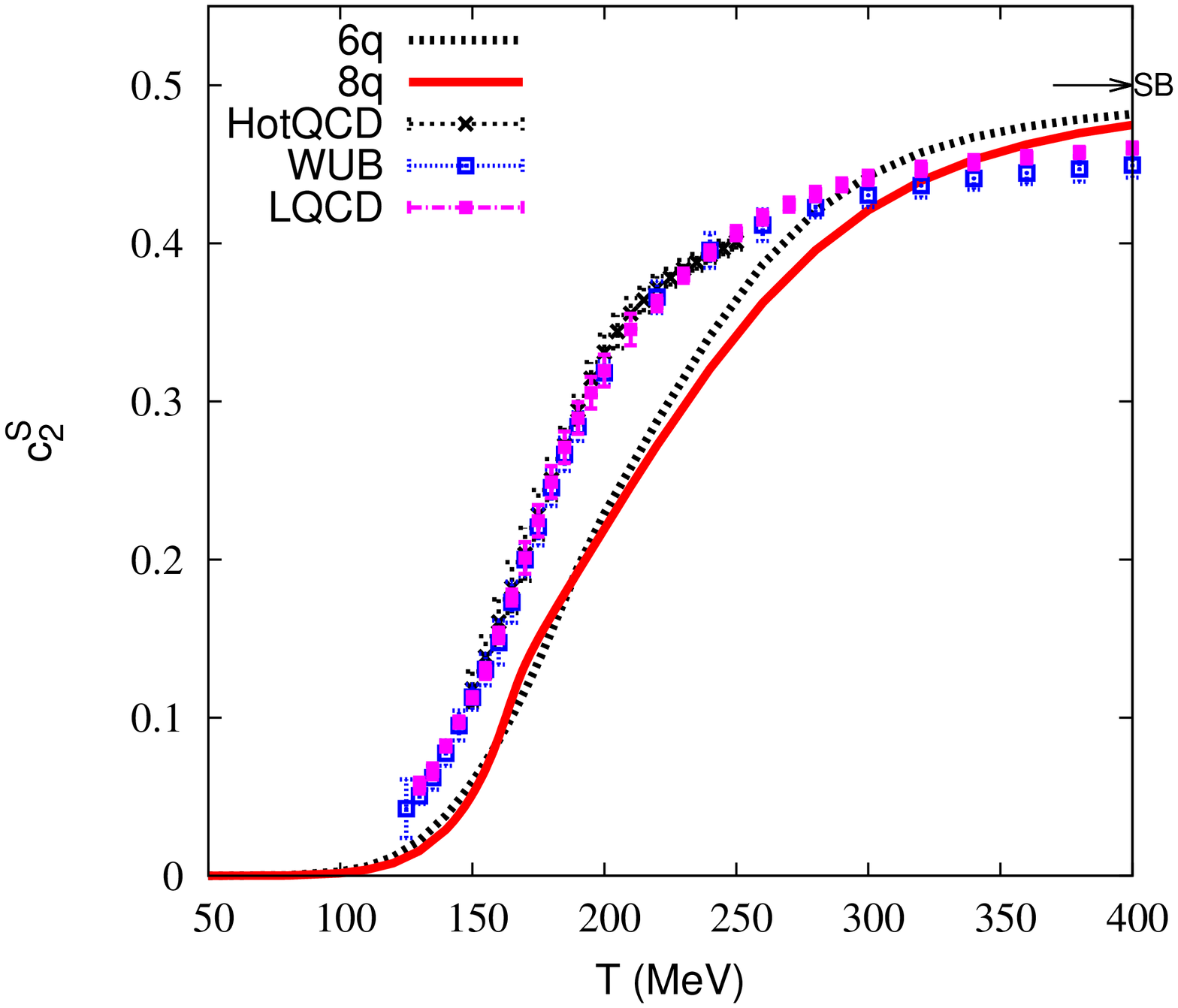}
\includegraphics[scale=0.38,angle=270]{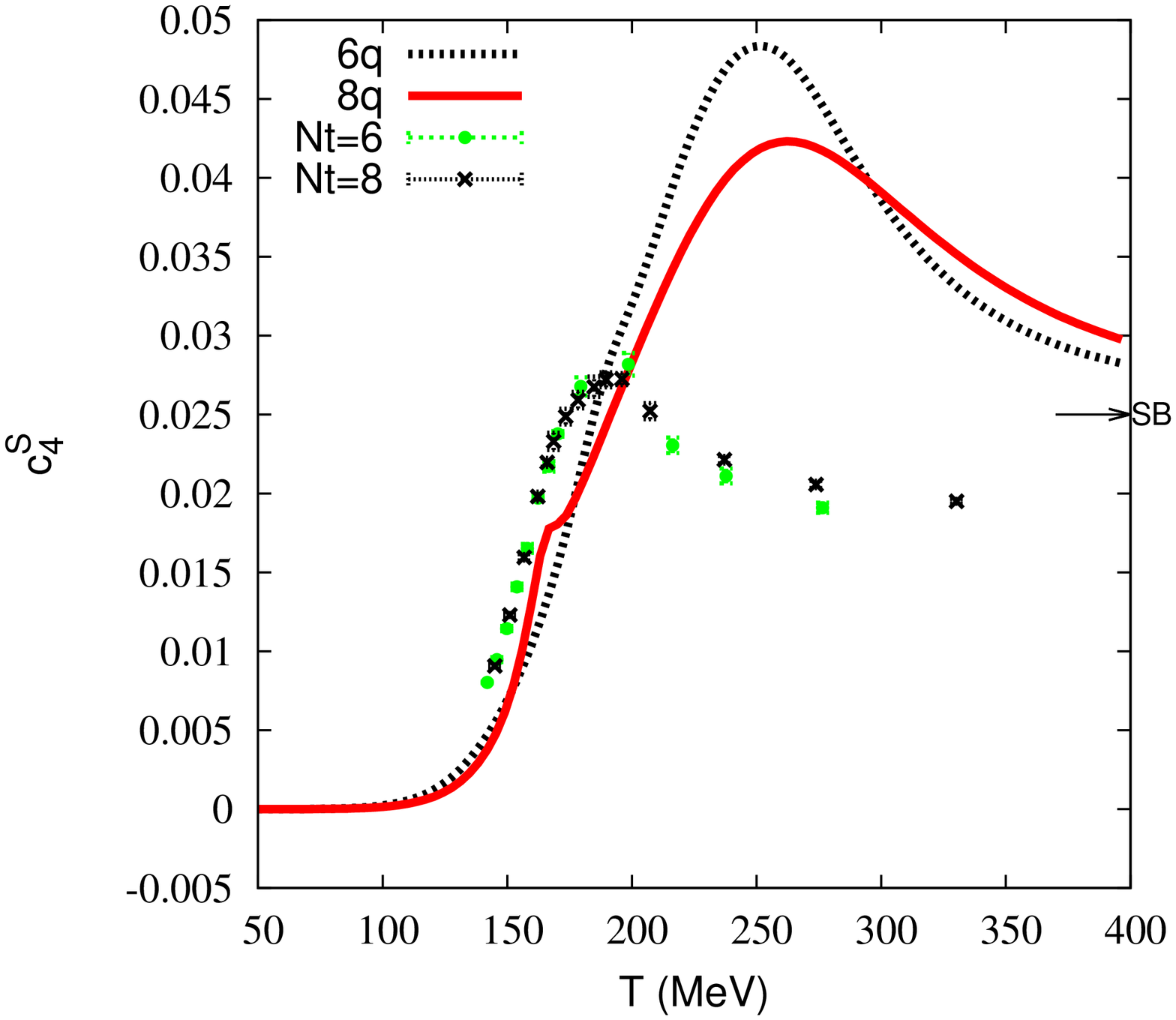}
}
\caption{(color online) Variation of $c_2^S$ and $c_4^S$ as functions of
temperature. The continuum extrapolated dataset for $c_2^S$ of HotQCD
and Wuppertal-Budapest (WUB) collaborations are taken respectively from
~\cite{BazavovPRD} and ~\cite{BorsanyiJHEP}, as well as the data from
~\cite{Bellwied2} denoted as LQCD. For $c_4^S$ HotQCD data for Nt=6 and 8
~\cite{BazavovPRL}.}
\label{fg.cS}
\end{figure}

\par
The variation of the charge susceptibilities with temperature are shown
in Fig.~\ref{fg.cQ}. The qualitative as well as quantitative comparison
between the two interaction models and the lattice QCD data are quite
similar to that discussed for the baryon number susceptibilities for
$T>T_c$. However we now find significant difference between PNJL and
lattice results for $c_2^Q$ below the crossover temperature $T_c$. The
lattice data is much larger than the model results. This seems to be
expected from our earlier discussions of discrepancies in speed of
sound. In the charge sector the dominant contributors are the light
hadrons, and these excitations are effectively absent in the present
form of the PNJL model.  Therefore though the baryon fluctuations are
well accounted for by the constituent quarks, proper considerations of
other hadronic degrees of freedom below $T_c$ is crucial to obtain the
charge fluctuations. 

\par
The temperature variation of the strangeness susceptibilities $c_2^S$
and $c_4^S$ are shown in Fig.~\ref{fg.cS}. Here also the quantitative
results of $c_2^S$ are found to be different between the model and lattice
QCD data up to $T_c$. Proper inclusion of the light strange hadrons
would be crucial in describing this region of
temperature~\cite{Swagato3}. Above $T_c$ the agreement is again much
better.  However for $c_4^S$ there is a large difference between the
PNJL model results and lattice data for $T>T_c$.  The maxima obtained in
the model is much larger, wider, as well as shifted towards higher
temperatures as compared to the lattice data.  As discussed by some of
us earlier in Ref.~\cite{Deb} this is due to the melting of the strange
quark condensate at higher temperatures in the PNJL model. This is
possibly an artefact of constraining the NJL model parameters to be
fixed at values obtained at zero temperature and chemical potentials. It
would be important to investigate the necessary changes in the quark
interactions in the NJL Lagrangian, but is beyond the scope of the
present work.

\begin{figure}[!htb]
{\includegraphics[scale=0.28,angle=270]{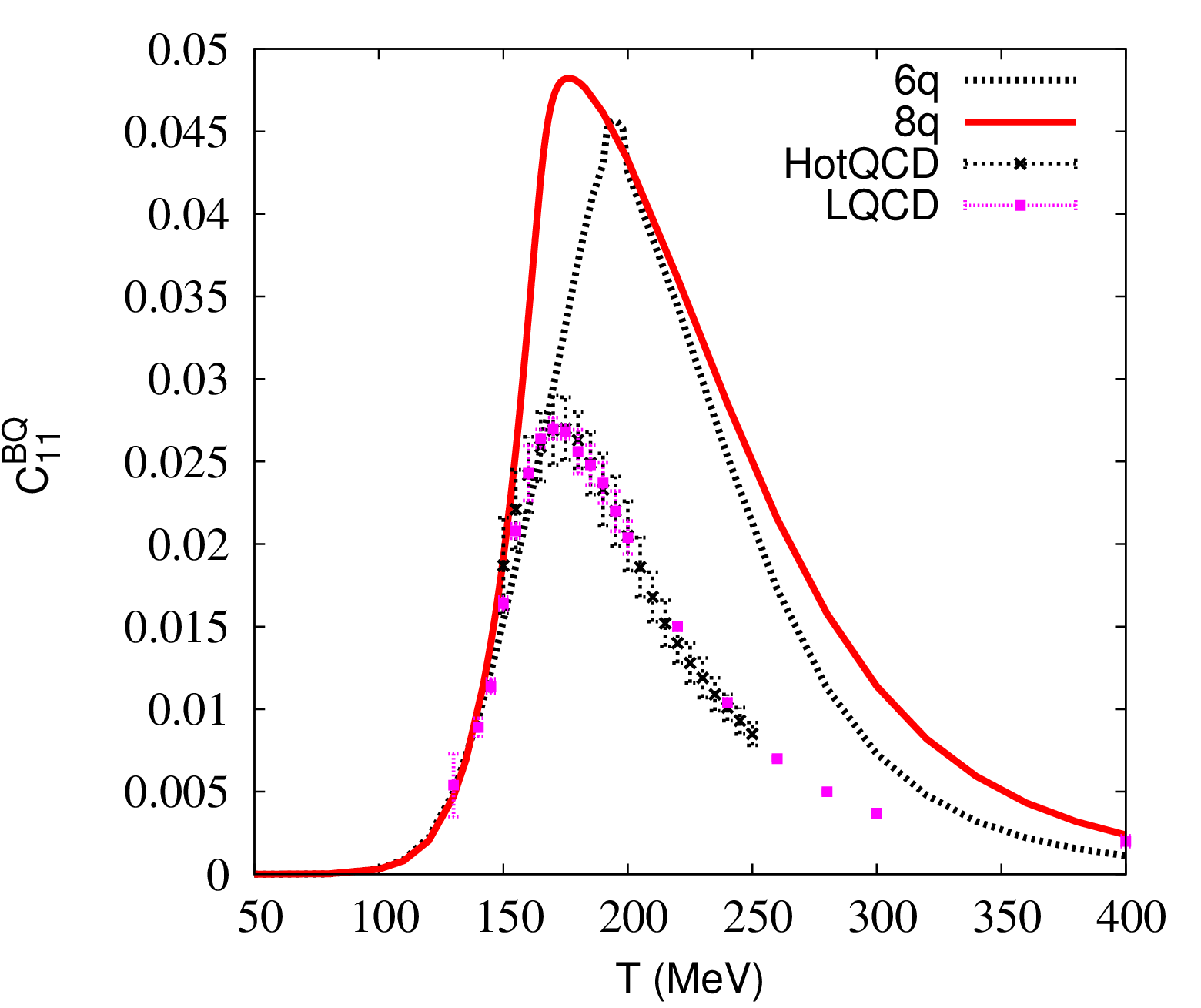}
\includegraphics[scale=0.28,angle=270]{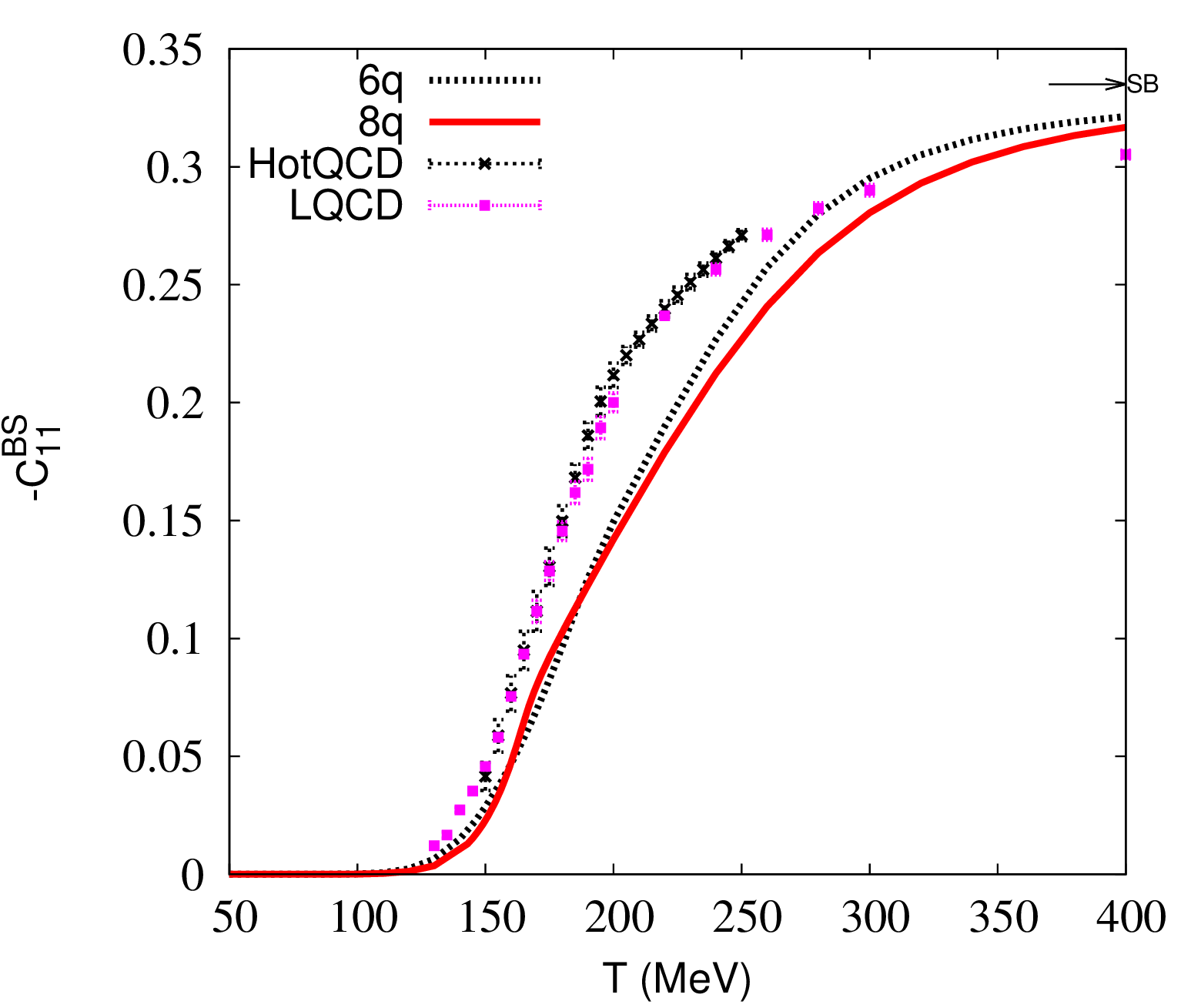}
\includegraphics[scale=0.28,angle=270]{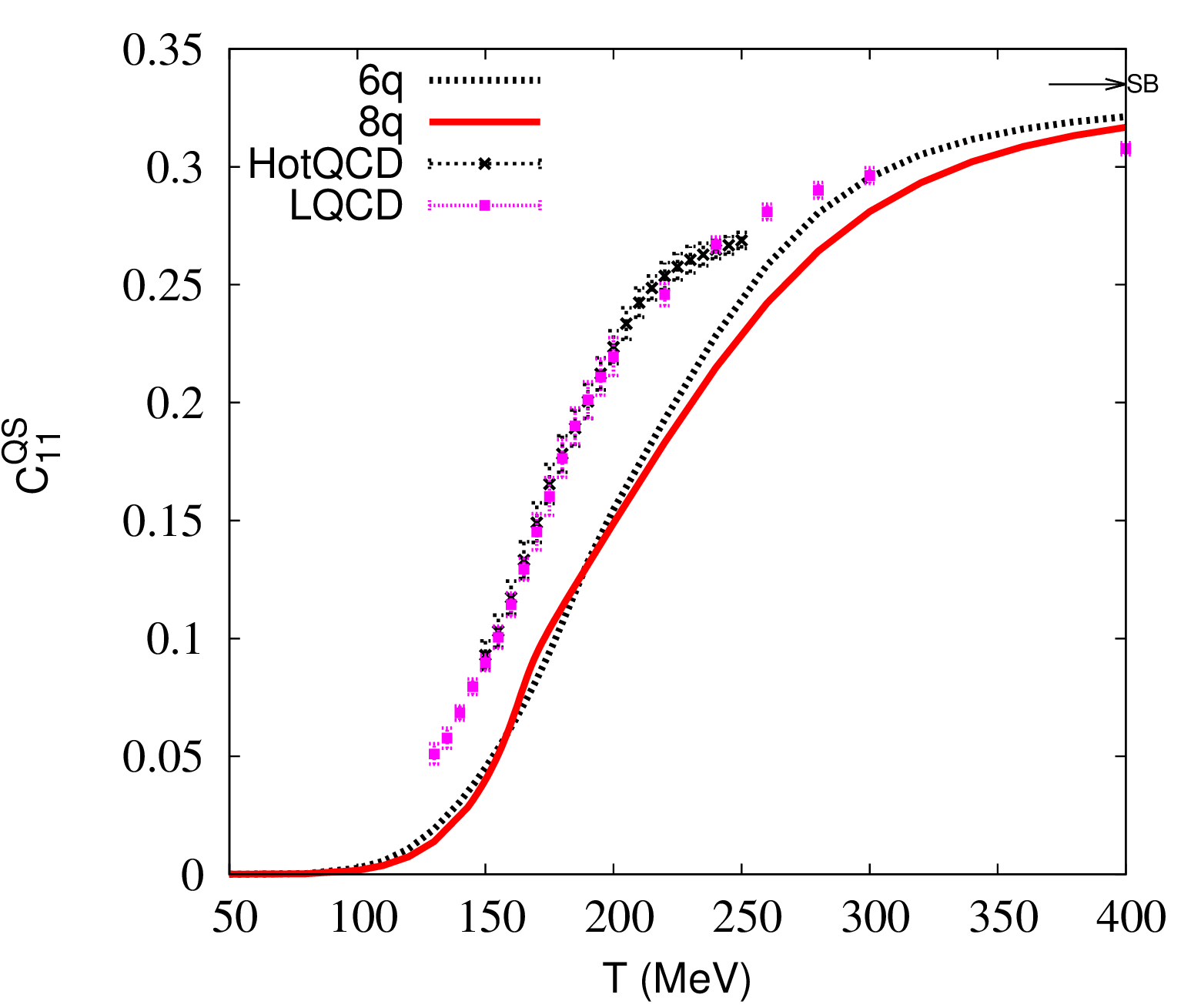}}
\caption{(color online) Leading order correlation coefficients as a
function of temperature.  HotQCD continuum dataset have been taken from
\cite{BazavovPRD}.Lattice data from \cite{Bellwied2} have denoted as LQCD.} 
\label{fg.corr11}
\end{figure}

\par
We now discuss the leading order correlations between the conserved
charges. These are shown as functions of temperature in
Fig.~\ref{fg.corr11}. The baryon number to electric charge (BQ)
correlation $c_{11}^{BQ}$ shows a hump around the crossover region and
vanishes for both low and high temperatures. In the hadronic phase the
baryon and electric charge are correlated because baryons have positive
electric charge and anti-baryons have negative electric charge. However
their masses being large, the correlations come out to be
insignificant. With increasing temperature however the correlation
becomes non-zero. On the other hand in the partonic phase, for the 2+1
flavor theory, there are three quarks with equal baryon number but
electric charge of $down$ and $strange$ quarks are together opposite of
that of the $up$ quark, implying that in this phase the BQ correlation is
zero. Thus we get the temperature variation of BQ correlation 
as shown in Fig.~\ref{fg.corr11}. We note that the BQ
correlation in the PNJL model is larger than that obtained in the
lattice QCD data.

\par
The baryon number to strangeness (BS) correlation $c_{11}^{BS}$ as well
as the electric charge to strangeness (QS) correlation $c_{11}^{QS}$
show a order parameter like behavior. This is because at low
temperatures they are suppressed due to large hadronic masses, and
eventually increases with increase in temperatures. For these two
correlations we note that PNJL model results are significantly lower
than the lattice QCD data.  This is similar to the behavior of the
second order strangeness susceptibility $c_2^S$, which should be as we
discuss below.  For these correlators we find the lattice results to be
larger than the PNJL results.

\par
\begin{figure}[!htb]
{\includegraphics[scale=0.28,angle=270]{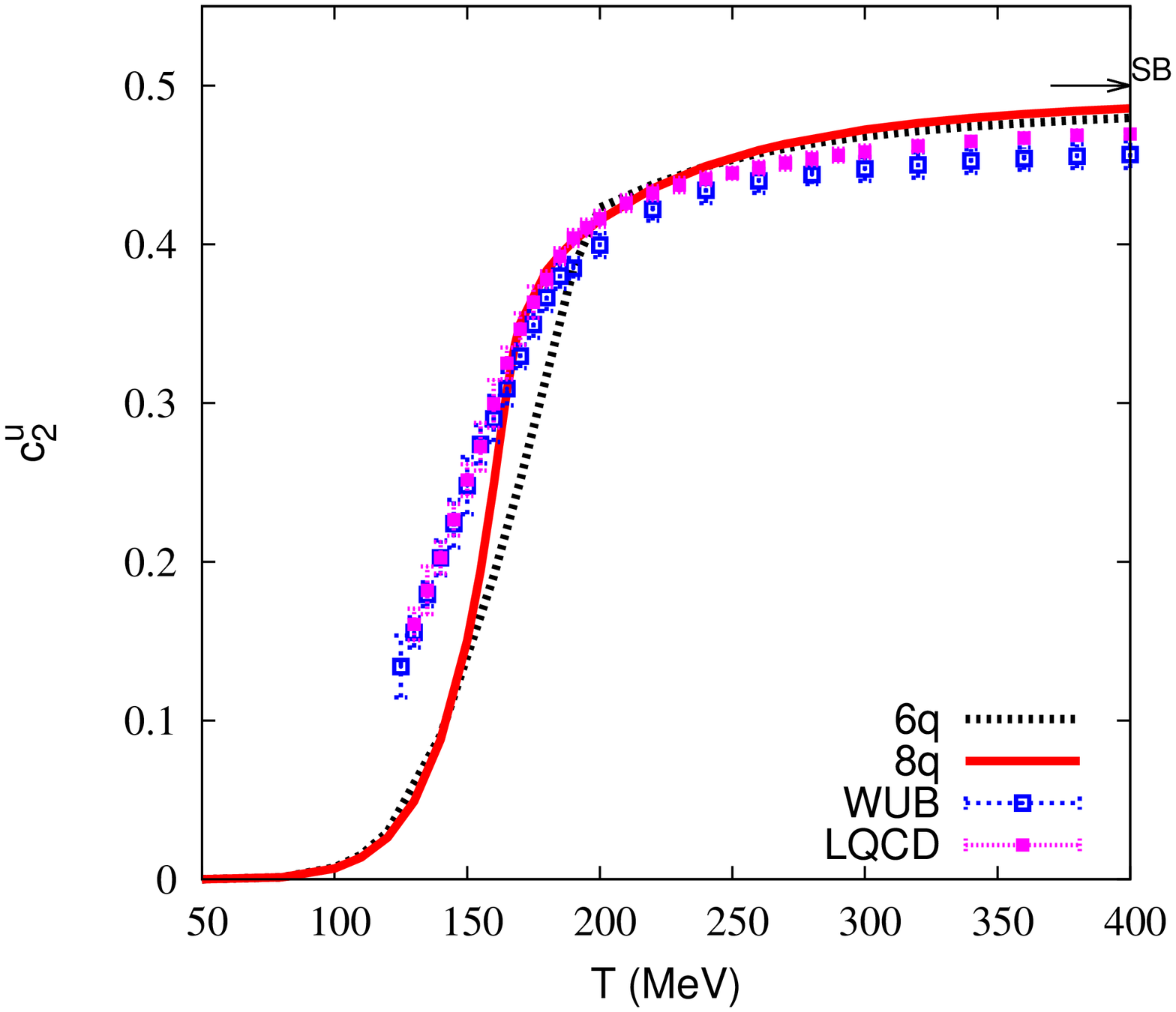}
\includegraphics[scale=0.28,angle=270]{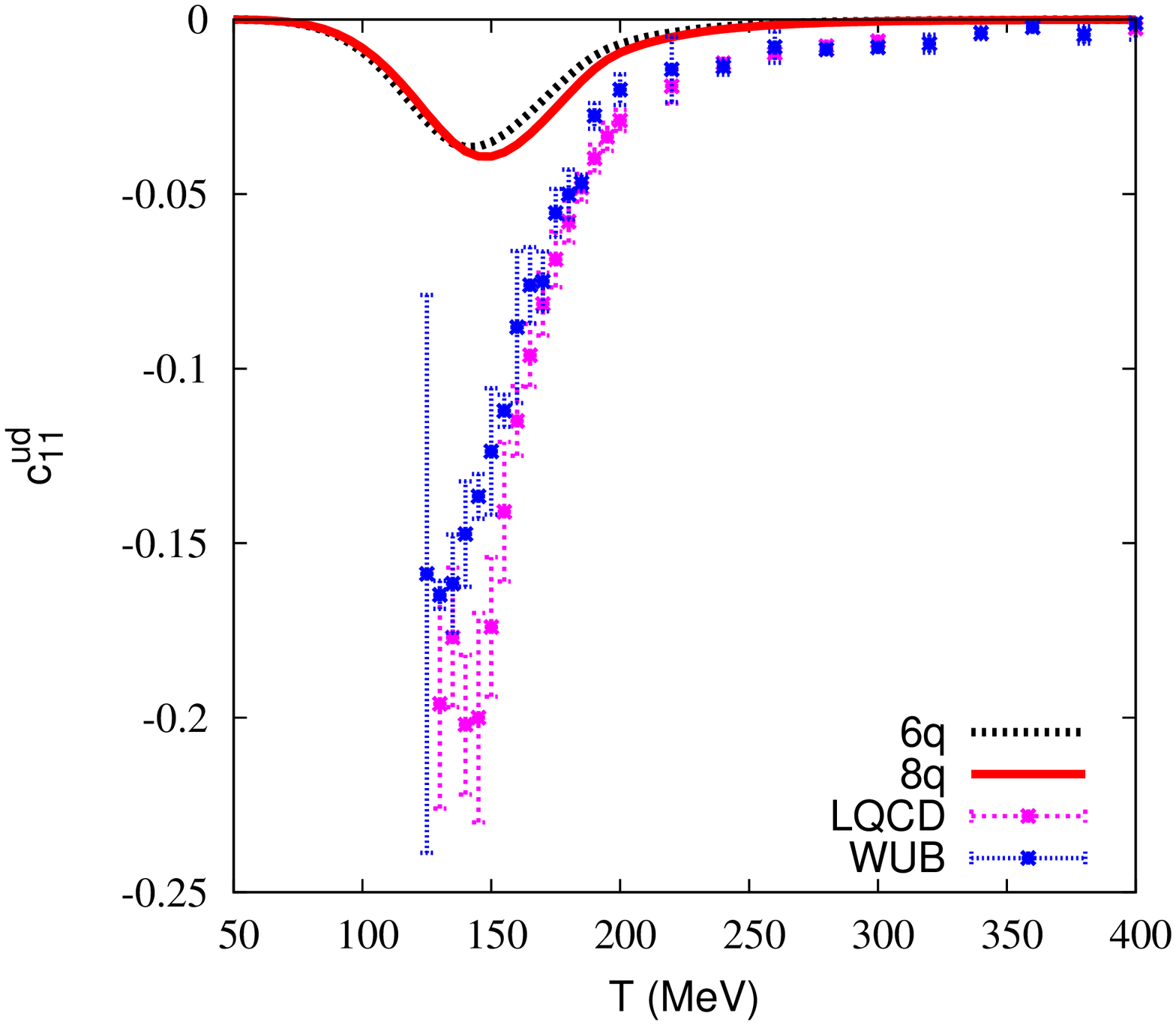}
\includegraphics[scale=0.28,angle=270]{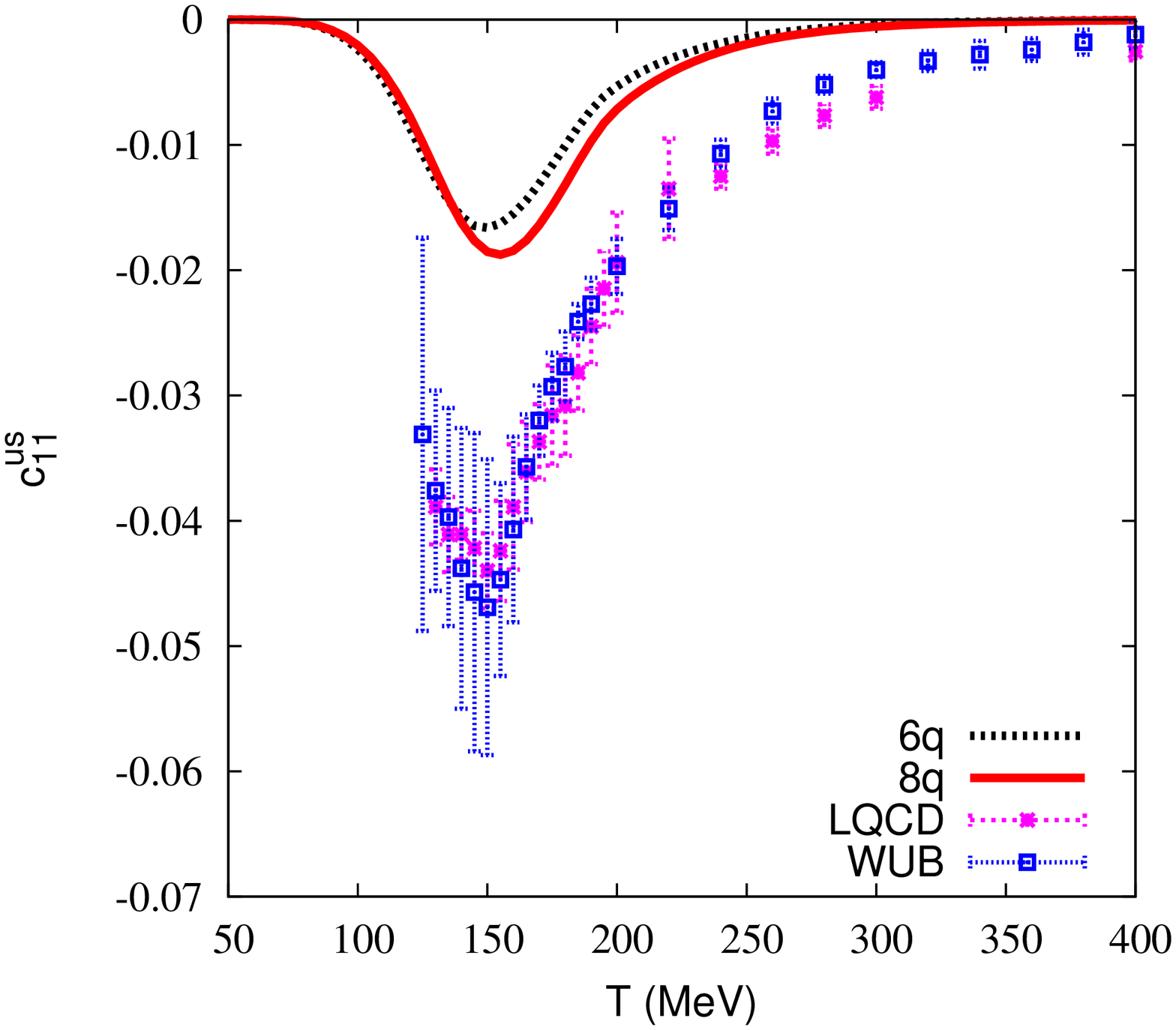}}
\caption{(color online) Leading order fluctuations and correlations
of quark flavors as a function of temperature.
Lattice continuum data for WUB are from ~\cite{BorsanyiJHEP}
and LQCD are from \cite{Bellwied2}.} 
\label{fg.flav}
\end{figure}

Now it seems strange that the correlators at the same order
have opposite behavior for $c_{11}^{BQ}$ versus $c_{11}^{BS}$ and
$c_{11}^{QS}$, when PNJL model is compared to the lattice QCD data. Let
us try to argue how this could naturally arise. For that we first
express the correlators in terms of the fluctuations and correlations in
terms of the flavor basis. The relations are given as,

\begin{eqnarray}
c_{11}^{BQ}&=&\frac{1}{9}\left(c_2^u-c_2^s+c_{11}^{ud}-c_{11}^{us}
\right),\\
c_{11}^{BS}&=&\frac{1}{3}\left(-c_2^s-2c_{11}^{us}\right),\\
c_{11}^{QS}&=&\frac{1}{3}\left(c_2^s-c_{11}^{us}\right),
\end{eqnarray}

\noindent
where $c_2^u=c_2^d$ and $c_2^s$ are the second order flavor
susceptibilities and $c_{11}^{ud}$ and $c_{11}^{us}$ are the second
order flavor correlations. We note that if we consider the flavor
correlators to be numerically much smaller than the flavor
susceptibilities one may again describe the observed behavior of the
correlators in Fig.~\ref{fg.corr11}. The up flavor diagonal and the
off-diagonal susceptibilities are presented in Fig.~\ref{fg.flav}.  The
strange flavor diagonal susceptibility is identical to the strangeness
diagonal susceptibility as shown in Fig.~\ref{fg.cS}.  While
$c_{11}^{BS}$ and $c_{11}^{QS}$ will inherit the order parameter like
behavior of $c_2^s$, $c_{11}^{BQ}$ will vary depending on the difference
between $c_2^u$ and $c_2^s$. This may explain the higher value obtained
in the PNJL model with respect to lattice QCD data. To see this we note
that in ~\cite{Raha} some of us discussed the variation of the baryon
number to isospin (BI) correlation
$c_{11}^{BI}=\frac{1}{6}\left(c_2^u-c_2^d\right)$ with different current
masses for the $up$ and $down$ quarks in a 2 flavor system.  For
identical light quark masses, $c_{11}^{BI}$ should be zero, but it
becomes non-zero when the current masses are different.  It was further
discussed that value of $c_{11}^{BI}$ is proportional to this mass
difference and for small quark masses it has a consistent scaling with
the amount of mass splitting. Here for $c_{11}^{BQ}$ a similar situation
arises due to the large strange quark current mass difference with that
of the light quarks.

\par

For the PNJL model we have considered the current quark masses as given
in Table ~\ref{tb.njlpara}.  For the lattice QCD data the bare quark
mass in physical units are found to have an average value of $m_s=81
\rm{MeV}$ (with a spread of 2 MeV), for the temperature range of the
data as obtained from Table XII of Ref.~\cite{Bazavov12}. This
difference in the bare masses may account for the difference in BQ
correlation between PNJL model and the lattice QCD results. A detailed
study in this direction will be presented elsewhere.

\par
The strange quark mass being smaller for the lattice data it is highly
conceivable that the second order susceptibilities are higher on the
lattice exactly as observed in the behavior of $c_{11}^{BS}$ and
$c_{11}^{QS}$. This would also partially be responsible for the large
difference of $c_2^S$ obtained in the PNJL model and on the lattice.  A
proper reparametrization of the NJL model with lower current mass for
the strange quark may therefore bridge the gap in the various
susceptibilities and correlations related to the strangeness sector and
will be addressed elsewhere.  It should also be noted that a further
suppression to the BQ correlation in the lattice data is due to a
significant contribution from the $ud$ correlation. The flavor
correlations in the PNJL model are quite suppressed compared to the
continuum lattice data, which is probably due to the lack of proper
considerations of the hadronic degrees of freedom.

\section{Conclusions}
\label{sc.conclude}

\par
QCD in the non-perturbative domain is best realized with lattice QCD
simulations which are however very costly. Simpler model approaches are
efficient in the extraction of the quantities of interest at arbitrary
values of external parameters like temperature, chemical potential etc.
which however needs to have reliability validated quantitatively. In
this work we discussed how far the PNJL model is suitable in describing
the thermodynamic properties of strongly interacting matter.  Recently,
lattice QCD simulations have been extrapolated to the continuum limit
and almost physical quark masses, obtaining a variety of interesting
information for a wide range of temperature. Therefore it seemed timely
that a reparametrization of the PNJL model be made to check if it
can satisfactorily predict various measured observables on the lattice.

\par
An important observation in the continuum extrapolated lattice results
is that the pressure of strongly interacting matter is significantly
below that of ideal gas of quarks and gluons even at reasonably large
temperatures. This implies that the gluon mediated interactions must be
strong even though the degrees of freedom may have changed from hadronic
to partonic ones. So we chose to reparametrize the Polyakov loop self
interactions in the PNJL model which is supposed to mimic the gluonic
effects. The NJL model parameters were set from hadronic properties at
zero temperature and chemical potentials. 

\par
We found excellent agreement of the equation of state in the PNJL model
with that of lattice QCD data in a wide range of temperatures. The
specific heat has a small peak in the model near the crossover in the
model. Though not a prominent peak but a hump is surely present in the
lattice QCD data. The speed of sound agrees with lattice data except for
$T<T_c$.

\par
The second and fourth order susceptibilities of the baryon number were
again found to be in reasonable quantitative agreement with the lattice
data. For the electric charge susceptibilities we found some
disagreement for $T<T_c$. The disagreement in this region for speed of
sound as well as susceptibilities could possibly be due to absence of
light hadrons in the present formulation of the PNJL model.

\par
Significant disagreement was observed for baryon-charge,
baryon-strangeness and charge-strangeness correlations. The values were
more in the PNJL model for the baryon-charge correlation and opposite
for the other correlators.  We argued that this could possibly due to
the difference in the bare strange quark masses used in the PNJL model
and the lattice formulations. With this argument the opposing
discrepancies in the correlators could also be explained. This could
also be partially responsible for the discrepancies in the strangeness
susceptibilities. The most significant disagreement is observed for the
fourth order susceptibility of strangeness for $T>T_c$. The slow melting
of the strange quark condensate seems to be a major cause for this
discrepancy.

\par
Thus even though the quantitative agreement of a variety of observables
in the PNJL model with the lattice QCD data was found to be encouraging,
certain differences still remain. A proper consideration of hadronic
excitations and reparametrization of the NJL part seems necessary. We
would like to address these issues elsewhere.

\section*{Acknowledgements}
The authors would like to thank Council for Scientific and Industrial
Research (CSIR), Department of Science and Technology (DST) 
and Alexander von Humboldt Foundation (AvH) for financial support
for funding this work. We thank Johannes Weber for useful suggestions.


\end{document}